\documentclass[lettersize,journal]{IEEEtran}

\ifCLASSINFOpdf
  \usepackage[pdftex]{graphicx}
\else
\fi
\usepackage{hyperref}
\usepackage{comment}
\usepackage[switch]{lineno}
\usepackage{enumitem}
\usepackage{graphicx}
\usepackage[normalem]{ulem}
\useunder{\uline}{\ul}{}
\begin{document}

%
\title{How are Drivers' Stress Levels and Emotions Associated with the Driving Context? \\ A Naturalistic Study}


\author{Arash~Tavakoli, Nathan Lai, Vahid Balali, and Arsalan Heydarian
\IEEEcompsocitemizethanks{\IEEEcompsocthanksitem Arash Tavakoli and Arsalan Heydarian (Corresponding Author) are with the Department
of Engineering Systems and Environment, University of Virginia,
VA, 22901.\protect
E-mail: ah6rx@virginia.edu
\IEEEcompsocthanksitem Nathan Lai is with the department of Computer Engineering and Computer Science, California State University, Long Beach,
CA, 90840.\protect
\IEEEcompsocthanksitem Vahid Balali is with the department of Civil Engineering and Construction Engineering Management, California State University, Long Beach,
CA, 90840.\protect
}}

\maketitle

\begin{abstract}
Understanding and mitigating drivers' negative emotions, stress levels, and anxiety is of high importance for decreasing accident rates, and enhancing road safety. While detecting drivers' stress and negative emotions can significantly help with this goal, understanding what might be associated with increases in drivers' negative emotions and high stress level, might better help with planning interventions. While studies have provided significant insight into detecting drivers' emotions and stress levels, not many studies focused on the reasons behind changes in stress levels and negative emotions. In this study, by using a naturalistic driving study database, we analyze the changes in the driving scene, including road objects and the dynamical relationship between the ego vehicle and the lead vehicle with respect to changes in drivers' psychophysiological metrics (i.e., heart rate (HR) and facial expressions). Our results indicate that different road objects might be associated with varying levels of increase in drivers' HR as well as different proportions of negative facial emotions detected through computer vision. Larger vehicles on the road, such as trucks and buses, are associated with the highest amount of increase in drivers' HR as well as negative emotions. Additionally, shorter distances and higher standard deviation in the distance to the lead vehicle are associated with a higher number of abrupt increases in drivers' HR, depicting a possible increase in stress level. Our finding indicates more positive emotions, lower facial engagement, and a lower abrupt increase in HR at a higher speed of driving, which often happens in highway environments. This research collectively shows that driving at higher speeds happening in highways by avoiding certain road objects might be a better fit for keeping drivers in a calmer, more positive state.

\end{abstract}



%
\IEEEpeerreviewmaketitle

\section{Introduction}
Designing human-centered vehicles require understanding how different factors in and out of the vehicle can affect a user's state, such as stress level, workload, and anxiety \cite{bustos2021predicting}. This is mostly due to the fact that drivers' decision-making and resulting behaviors are affected by their emotional and cognitive states \cite{taubman2012effects,roidl2014emotional}, which ultimately have severe impacts on driving safety. For instance, recent studies through naturalistic environments suggest that negative emotions, higher stress levels, and cognitive load may increase the probability of driving accidents and certain driving behaviors such as risky driving \cite{dingus2016driver,sani2017aggression,shukri2022theory}. While multiple studies are targeting emotion detection in-cabin, not that many are focused on understanding what parts of the environment (in-cabin and on the road) might be associated with changes in driver's states such as inducing negative emotions and higher cognitive load \cite{zepf2019towards}. In line with addressing this gap, recent research has started to analyze the driving environment together with subjective self-reports of driver's stress to find possible correlations between environmental attributes and driver's stress level. For instance, recent studies found that certain road objects such as bigger vehicles (e.g., trucks), road users (e.g., cyclists), and infrastructural elements (e.g., intersections), as well as in-cabin situations (e.g., working with the center stack), are highly associated with higher subjective stress levels \cite{zepf2019towards,dittrich2021drivers,bustos2021predicting}. In addition to road objects, some research studies showed that car-following distance and behavior could also affect drivers' psychological states \cite{nacpil2021application}. For instance, \cite{zheng2015biosignal} showed that within a simulated platooning scenario, drivers' mental stress measured through biosignals increased as the distance to the lead vehicle decreased. Another recent study performed in a driving simulator found out that drivers' workload was higher during a shorter time headway \cite{carfollow}.  

Understanding the aforementioned reasons behind drivers' state including emotions, workload, and stress induction, can help mitigate them in driving by choosing less stressful routes \cite{tavakoli2021leveraging}, personalizing car-following distance \cite{nacpil2021application}, or by providing interventions (e.g., listening to music) \cite{fakhrhosseini2019angry,niu2020music}. While previous studies provided significant evidence on the correlation between drivers' subjective measures of stress and environmental attributes, it is still difficult to apply these findings in real-world conditions as most of these studies are centered on subjective measurements. Recent developments in ubiquitous computing such as smartwatches are facilitating their applications in detecting unhealthy states of the users such as abrupt changes in anxiety, stress level, and experiencing negative emotions \cite{wang2021personalized,tavakoli2021harmony}. Using ubiquitous computing devices, studies have found strong correlations between human psychophysiological measures (e.g., HR and skin conductance) and stress level and work load, and more specifically in driving, studies show that increase in human HR might be correlated with stressful experiences \cite{napoletano2018combining,chung2019methods,lohani2019review,laora2022,tavakoli2021harmony}. 

In addition to smartwatches, advancements in computer vision techniques have made it viable to detect certain objects fully automatically without any manual annotation. For example, recent developments in end-to-end object detection algorithms (e.g., see MASK RCNN \cite{matterport_maskrcnn_2017}) together with high-quality datasets (e.g., see COCO \cite{lin2014microsoft}) have made it possible to detect many road objects such as signs, road users, and infrastructural elements. Even in the case of not having access to off-the-shelf models, current computer vision frameworks are making it easier to build newer models for different road objects. Coupling smartwatches with features extracted from videos can help with finding possible associations between the presence of road objects and drivers' stress changes, objectively. This information can then be leveraged to enhance the driving experience and mitigate possible faulty decisions as a result of being under stress. In the context of automated driving, these methods can help with take-over control of the vehicle in a much more efficient and faster manner. 

In this paper, we take an exploratory approach to understanding the relationship between changes in drivers' stress levels and emotions in real-world driving context by using multi-modal naturalistic driving data, which includes drivers' psychophysiological measures, and behavioral metrics (vehicle speed) as well as outside-cabin environment videos. Based on a naturalistic driving dataset, namely HARMONY \cite{tavakoli2021harmony}, we first retrieve drivers' facial expressions as well as abrupt increases in their HR, which might be indicative of increases in stress level \cite{tavakoli2021harmony}. We detect abrupt increases by using a change point detector based on Barry and Hartigan's method \cite{barry1993bayesian}. We analyze the driving scene retrieved from the video recordings by (1) detecting road objects and (2) estimating the relative distance to the lead vehicle. By analyzing the co-occurrence of the abrupt increases in drivers' HR and the presence of lead vehicles, we find that different road objects might be associated with varying levels of increases in drivers' HR, indicating different stress levels as well as different fractions of negative facial emotions. Our results indicate that larger vehicles on the road, such as trucks and buses, might be associated with the highest amount of increase in drivers' HR as well as negative emotions. Additionally, our findings suggest that shorter distances and higher standard deviations in the car-following distance, might be associated with a higher number of abrupt increases in drivers' HR, indicating a higher stress level. Moreover, our findings indicate more positive emotions, less facial engagement, and a lower number of abrupt changes in HR at a higher speed of driving.

\section{Background}
Studies in the past have provided significant information on the interplay of drivers' unhealthy states (e.g., emotions, stress level, anxiety, and cognitive load) and the environmental attributes. These studies have shown that environmental attributes such as in-cabin situation, road types, road users, weather, and in-cabin conditions can affect how a driver feels and can result in affecting their driving performance and behaviors both in semi-automated (e.g., take over control) and manual driving (e.g., lane keeping) \cite{zepf2019towards,dittrich2021drivers,bustos2021predicting}. As this paper is mostly centered on drivers' emotions and stress levels, we first review these concepts from a psychological point of view. Then, we discuss how these concepts have been applied to evaluate drivers' state in different roadway conditions in driving research. 

Understanding emotion and its applications have been one of the main topics of psychology, philosophy, neuroscience, artificial intelligence and computer-human interaction. Psychology literature provides different theories of emotion, such as the categorical and dimensional emotion theories. The categorical, sometimes referred to as the basic emotion theory, posits that there is a specific limited number of emotions that are basic psychological and biological concepts and cannot be divided into more basic ingredients. Although there is not a consensus on the number of the basic emotions and as to which emotion is a basic emotion, but there seems to be an agreement on the definition of the basic emotion. These emotions are distinct in their recurring fixed patterns of neural and bodily expressed components and physiological and behavioral signatures such as variation in heart rate, and facial muscle movements \cite{izard2009emotion,tracy2011four,ekman2011meant,barnard2016anxiety} which is in response to a stimulus. Different psychologists proposed a different number of basic emotions and accounted various emotions as basic. For instance, Izard proposed six basic emotions of happiness, sadness, fear, anger, disgust, and interest, while Ekman proposed the basic emotions to be happiness, sadness, fear, anger, disgust, contempt, and surprise \cite{tracy2011four}. 

Dimensional emotion theory proposes that emotions can be represented with numerical values in multiple dimensions. One of the famous dimensional emotion theories is Russell's dimensional emotion model, where emotions are represented by their valence and arousal in a two-dimension format \cite{russell1980circumplex}. Valence refers to the level of positivity and negativity of emotion, whereas arousal refers to the level of activation in each emotion. In this model, an emotion such as ``excited'' has relatively high positive valence and high arousal, whereas an emotion such as ``bored'' has a negative valence with very low arousal. 

Driver stress is defined as the process of facing a situation where the perceived demand, mostly defined based on the previous experiences, internal body sensations, and external stimuli, is higher than the available resources \cite{francis2018embodied}. Stress can happen at different time scales where short-term stress is referred to as acute stress, in contrast to long-term stress, which is referred to as chronic stress \cite{francis2018embodied}. Multiple studies in driving research have attempted to detect changes in emotion and stress level through measuring human physiological metrics such as facial expressions, cardiac measures, and skin temperature and conductance \cite{chesnut2021stress,giannakakis2017stress,kim2018stress,tavakoli2020personalized,du2020psychophysiological}. These studies are mostly based on the assumption that changes in human physiology follow similar patterns within each specific emotional state. For example, studies show that increases in human HR might be correlated with an increase in stress level and negative emotions \cite{kim2018stress}. Additionally, studies have shown that the movement of facial muscles within each emotion category might follow specific patterns, where computer vision applications can be leveraged to detect emotions from the facial expressions \cite{mcduff2016affdex}. In a similar approach, certain patterns can be recognized while experiencing stress through machine learning applications \cite{giannakakis2017stress}. 

While multiple studies focused on detecting unhealthy states (e.g., stress level), not that many studies have analyzed the reason behind the elicitation of each state \cite{scott2018qualitative,zepf2019towards}. Understanding emotion triggers is of high importance as it helps with planning for interventions in driving, which can then help mitigate the effect of negative emotions and stress levels on drivers' performance, decision making, and take-over control. Mesken et al. analyzed three of the drivers' emotions (anxiety, happiness, and anger) by using an instrumented vehicle monitored by an experimenter in the vehicle \cite{mesken2007frequency}. The authors monitored 44 drivers'  speed, videos, and HR in an on-road controlled study. The participants were asked to verbally talk about their emotions as they faced any situation in driving. They found out that the emotion with the highest frequency was anxiety which was followed by anger and happiness. They identified that emotions were related to traffic events, such as driver's anger was associated with driving events that might affect their progress, while anxiety was related to driving events affecting safety. Additionally, the authors report an increase in HR associated with anxiety situations \cite{mesken2007frequency}. Roild et al. analyzed the responses of drivers' regarding the emotions that they experienced through a short survey \cite{roidl2013emotional}. In their study, the authors asked participants to rate their daily emotions in driving through an online questionnaire. Authors found out that drivers' anger, anxiety, and positive emotions were strongly related to situational factors \cite{roidl2013emotional}. Their results also point out that higher task demands are correlated with higher negative emotions \cite{roidl2013emotional}. Later, a study by \cite{zepf2019towards} monitored 33 drivers for a duration of 50 minutes through an on-road controlled study without an experimenter being present in the car. The authors also asked drivers to talk about their emotions as they faced them during the driving scenario. The authors analyzed 531 self-reports of drivers' voice recordings and provided four main categories of emotional triggers. The main categories included traffic \& driving task, environment, HCI \& navigation, and vehicle and equipment, which involved a few subcategories such as weather, other road users' behavior, and road designs \cite{zepf2019towards}. Another study performed by \cite{tavakoli2020personalized} found out that drivers' HR was lower in highways versus cities, clear versus adverse weather, and being with a passenger versus being alone. 

Another study by \cite{dittrich2021drivers} performed a similar analysis by monitoring 34 drivers' emotions with a focus on spatiotemporal triggers of emotions within an urban environment and found out that the main hotspot of emotional triggers are intersections. Additionally, they found out that environmental attributes such as other road users' behaviors and traffic lights had a higher fraction of negative emotions as compared to positive emotions within the self-reported stress. In their study, the authors point out that within the urban environment, highways are associated with the least stress level. While most of these studies were centered on self-reports, another study performed by \cite{tavakoli2021harmony} found out that different characteristics of the road environment might be associated with increases in drivers' HR. In their study, the authors found out that being followed by a vehicle, following a lead vehicle too closely, arriving at an intersection, and performing secondary tasks might be associated with abrupt increases in drivers' HR, possibly showing stress and negative emotions. Another study performed by \cite{tavakoli2022multimodal} analyzed drivers' psychophysiological measures (HR and gaze entropy) through a naturalistic study. In their study authors found out that drivers' had higher fraction of normal HR patterns as well as lower gaze entropy pattern within highway driving. Additionally, they found out that a more conservative driving style with close to zero acceleration was accompanied by lower fraction of abnormal HR and gaze entropy patterns. Lastly, a study by \cite{bustos2021predicting} attempted to predict self-reports of stress by solely relying on the snapshots of the visual scene. Based on using a convolutional neural network, the authors were able to predict the self-report associated with each visual scene in their database with an accuracy of 72 \%. Authors also pointed out that certain objects in the visual scene are correlated with higher subjective stress levels, such as traffic signals, bigger vehicles, and the presence of riders. Lastly, it should also be noted that previous studies in this area were mostly conducted through on-road controlled studies as well as using a driving simulator.  

Previous studies identify two major points regarding drivers' stress level and emotions. First, drivers' stress level and emotions are affected by the driving environment such as certain objects in the visual scene (e.g., presence of other road users, especially vulnerable road users, intersections and traffic lights), as well as lead vehicles and other traffic signs. Secondly, studies show that higher stress level in general is also correlated with increases in HR. Based on the previous literature we hypothesize the following:

The hypotheses for this paper are as follows:
\begin{enumerate}
    \item The attributes of changes in drivers' HR, valence, and facial engagement are significantly different across different environmental events including facing bigger vehicles, passing by intersections, presence of pedestrians, cyclists, and traffic signs. 
    \item Decrease in the car-following distance is correlated with higher levels of stress, which in turn is accompanied by increase in drivers' HR. 
\end{enumerate}

\section{Methodology}
This section is divided into multiple subsections describing the dataset (section \ref{sec:dataset}), and the detection of environmental perturbation (section \ref{sec:pert-detect}), changes in drivers' HR (section \ref{sec:hr-bcp}), drivers' facial emotions (section\ref{sec:affectiva}), and the distance to the lead vehicle (section \ref{sec:distance}). In these sections we further expand on how different factors are analyzed.

\subsection{Dataset}\label{sec:dataset}
\begin{figure*}
  \centering
  \frame{\includegraphics[width=\linewidth]{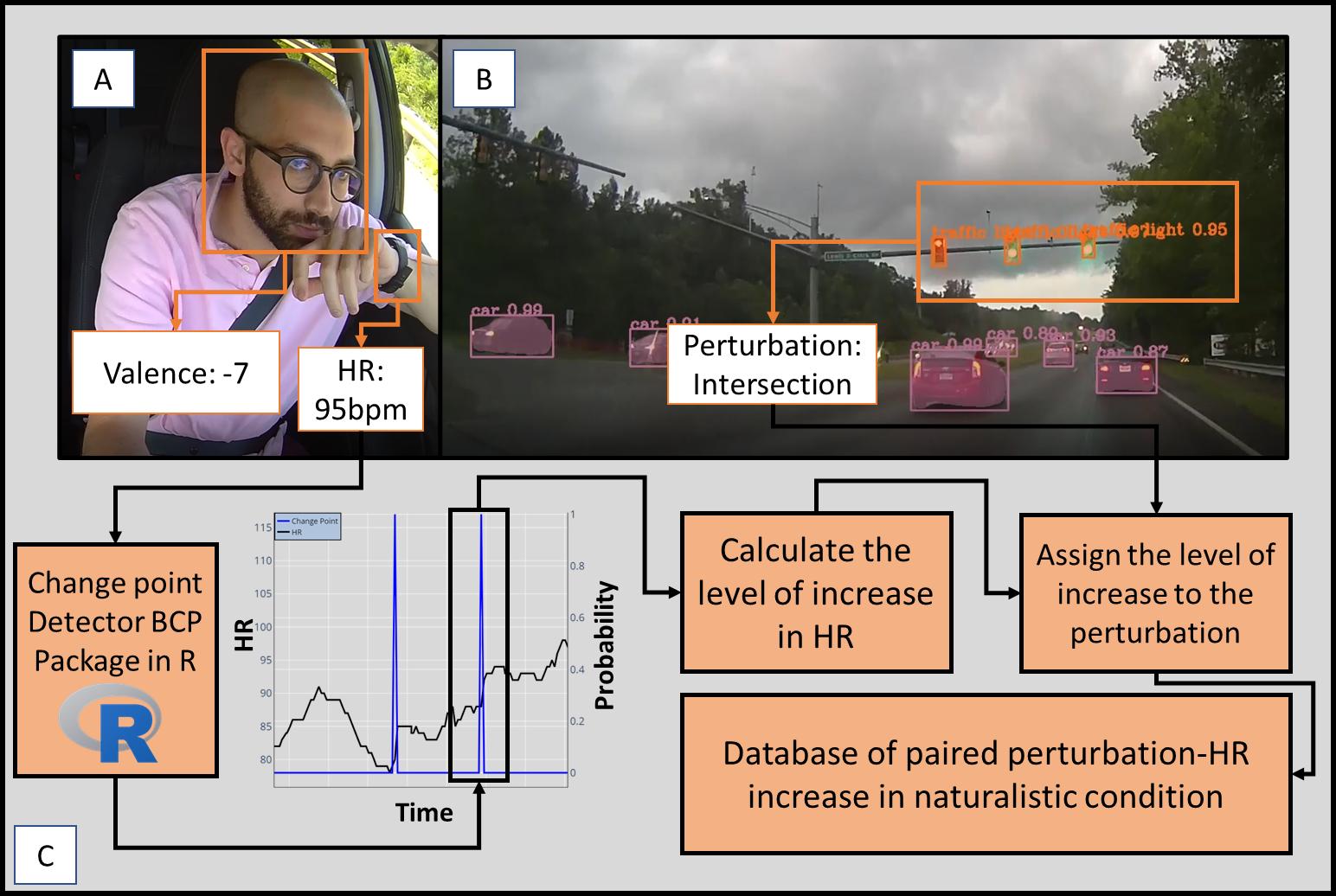}}
  \caption{A general view of the data from both in-cabin (A) and on road (B), as well as the application of change point detector for finding moments of high stress in drivers' HR (C).}
  \label{fig:framework}
\end{figure*}

The dataset for this study is provided by HARMONY, a human-centered multimodal study in the wild \cite{tavakoli2021harmony}. This dataset includes driving as well as human sensing data from 22 participants. The dataset is collected in a naturalistic fashion where each participant is provided with a camera and a smartwatch. The participants were asked to drive as they normally would in their daily lives. The camera recorded both in-cabin and outdoor environmental conditions. Additionally, the smartwatch collected the driver's HR, and acceleration, location, and environmental features such as noise and light level in-cabin. To this end, we collected and analyzed the data from 15 participants. A sample of the data is available online at \cite{Harmonydata}. Figure \ref{fig:framework} - A and B shows a sample of the data from both in-cabin and on-road situation points of view.

\subsection{Perturbation Detection}\label{sec:pert-detect}
In order to simplify the process of detection and analysis, this section is mainly focused on seven major environmental perturbations that were previously mentioned in driving research as stress-associated objects on the road \cite{zepf2019towards,dittrich2021drivers,bustos2021predicting}. These categories include the presence of speed limit signs, stop signs, intersections (i.e., traffic signals), big vehicles such as trucks and buses, riders such as bicycles and motorcycles, and pedestrians. Note that any other category can also be added to the analysis hereafter. However, we only focus on a set of perturbations that were already shown to affect subjective stress levels.

\subsubsection{Detection of Truck, Bus, Motorcycle, Bicycle, Traffic Signal, and Pedestrian}
In order to detect truck, bus, motorcycle, bicycle, traffic signal, and pedestrian, we used an off-the-shelf computer vision algorithm namely MASK RCNN. In this section, we used a pretrained model of the MASK RCNN algorithm \cite{matterport_maskrcnn_2017} that was trained on the Common Objects in Context (COCO) dataset \cite{lin2014microsoft}. A sample of the detection can be seen on figure \ref{fig:framework} - B.

\subsubsection{Detection of Stop Signs and Speed Limit}
In order to detect stop and speed limit signs within the pool of collected videos, we take advantage of the recent Computer Vision (CV) applications in sign detection. Current state-of-the-art CV algorithms (e.g., MASK R-CNN \cite{he2017mask}) that are trained on large datasets (e.g., COCO \cite{lin2014microsoft}) are capable of detecting stop signs. However, contrary to the test set provided by these algorithms, once applying them to our real-world videos collected in HARMONY, they often detect any sort of traffic sign as a stop sign, which increases the false positive rate. Additionally, if a stop sign is not facing the driver, it will still be detected, which is not applicable to our case as we are interested in stop signs that might be associated with a change in a driver's state. 

To overcome these issues, we retrained an object detection model on a stop and speed limit sign dataset, which was created by merging three external sign datasets. The model and code for this section are available through our GitHub \cite{githubLai}. We use transfer learning to retrain a model for detecting stop signs. In this regard, we take advantage of the YOLO V5 model, which is a recent modification of a deep learning object detection architecture, namely YOLO \cite{redmon2015you}. 

\paragraph{YOLO V5}
This model reimagines object detection as a regression problem and is inspired by the human visual system. YOLO is based on simultaneously predicting bounding boxes and the probability of each label associated with them. In this regard, YOLO can see the whole image at once and predict for the whole image rather than each individual bounding box, which results in less number of false positives \cite{redmon2015you}. YOLO architecture has 24 convolutional layers followed by two fully connected layers, as shown in detail in \cite{redmon2015you}. YOLO initially was suffering from different limitations, such as struggling with small objects in the visual scene. Different modifications were then added to the base YOLO, which resulted in YOLO versions 2 to 5. For this work, we focus on the most recent version of YOLO, which is YOLO V5 introduced by \cite{jocher2020yolov5,glenn_jocher_2021_5563715}. 

\paragraph{Sign Detection Datasets}
For this paper, we first trained YOLO V5 on a dataset of stop and speed limit signs. As mentioned previously, the dataset was created by merging three datasets of Laboratory for Intelligent and Safe Automobiles (LISA) \cite{jensen2016vision}, Common Objects in Context (COCO) \cite{lin2014microsoft}, and the dataset provided in  \cite{balali2016evaluation}. While both the LISA and Balali et al Sign datasets are only focused on traffic signs (e.g., stop, warning, and yield signs), the COCO dataset includes many road objects such as trucks, sedans, motorcycles, bicycles, and traffic signals. From combining the three datasets, 8,042 images were used as training which comprised 4,154 images of stop and speed limit signs. Within the stop sign and speed limit pool of images, 2,420 images were from the LISA dataset (1,291 stop signs and 1,129 speed limits), and 1,734 stop signs from the COCO dataset. Additionally, the training set included a total number of 3,888 negatives (none of the stop or speed limit signs) images, which comprised 250 negative images from the Balali et al. sign dataset, as well as 2,638 negative images from the LISA dataset. Lastly, 721 images were used as a test set, in which 652 images were from the LISA dataset and 69 images were from the COCO dataset (Table \ref{tab:stop_det}). 

\begin{table}[]
\caption{A summary of the dataset used for stop and speed limit sign detection}
\label{tab:stop_det}
\resizebox{0.49\textwidth}{!}{%
\begin{tabular}{l|lll}
                 & LISA  & COCO  & Balali et al. \\ \hline
Stop/Speed Limit & 2,420 & 1,734 & -      \\
Negative         & 2,638 & -     & 250   
\end{tabular}%
}
\end{table}

\begin{figure}
  \centering
  \frame{\includegraphics[width=\linewidth]{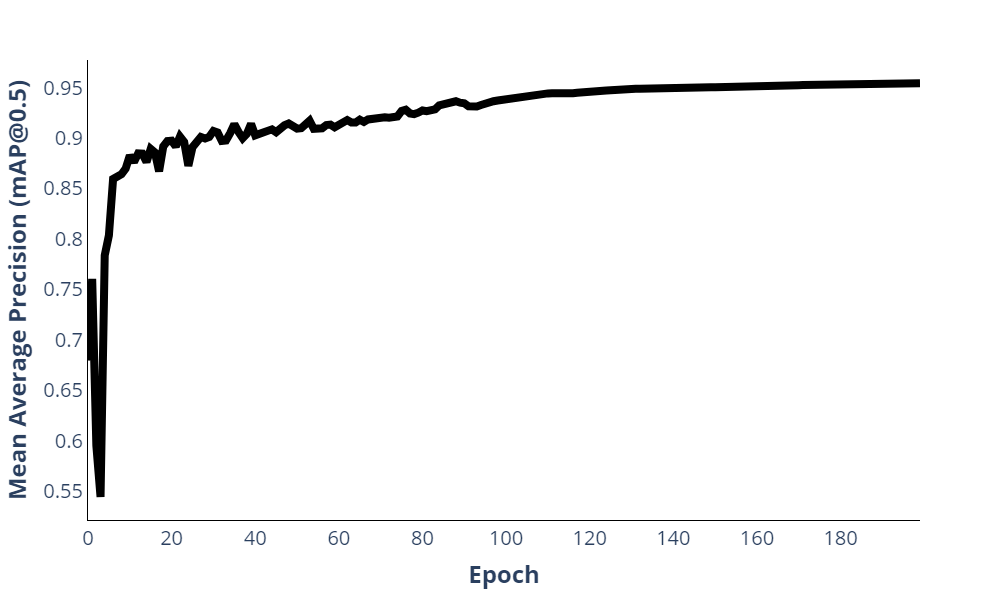}}
  \caption{The mean average precision of training yolov5 for stop sign and speed limit dataset }
  \label{fig:yolo_train}
\end{figure}

Utilizing YOLO V5 model, we trained the base model for 200 epochs with the default hyper-parameters. Figure \ref{fig:yolo_train} shows the mean average precision at 0.5 (mAP@0.5) for the stop sign and speed limit dataset. 

\paragraph{Optical Character Recognition (OCR)}
In order to further enhance the accuracy and to differentiate between speed limit signs, we integrated an Optical Character Recognition (OCR) \cite{nguyen2021survey} to detect signs that only have a specific text written on them. Using this method, we separate the signs that are not showing ``STOP'' or a number such as ``25'' on them. OCR is referred to the transforming of images into printed text. In order to apply OCR, we have tested the PyTesseract \cite{pytesseract}, and EasyOCR \cite{EasyOCR} packages. Our initial testing on the two packages showed that EasyOCR is slower but much more accurate in the images retrieved from real videos collected through HARMONY \cite{tavakoli2021harmony}. Thus we continue with the EasyOCR package. Every detection of the speed limit or stop sign is fed into the EasyOCR package. In the case of not detecting any character, it will be automatically removed from the detection. In the case of detecting ``Speed Limit``, we will then also find the number showing the limit. 


\subsection{Detecting Abrupt Increases in Drivers' HR}\label{sec:hr-bcp}
In order to detect the abrupt increases in drivers' HR, which, as mentioned previously, is correlated with increases in drivers' stress level, we take advantage of a change point detector. Due to motion artifacts introduced through different movements of drivers' hands, momentarily peaks can exist in the HR data, which is not of interest to our analysis. We are rather interested in detecting a change in the underlying distribution of the HR data. For this matter, we take advantage of a Bayesian Change Point (BCP) detector. BCP allows for easy quantification of uncertainty and integration of priors. Other studies have also mentioned the utility of BCP in detecting changes in data from different fields such as health \cite{malladi2013online}, transportation engineering \cite{tavakoli2021harmony,guo2021benchmarking,guo2021orclsim}, and behavioral science \cite{kumarlever,dong2021detection}. We leverage Barry and Hartigan's \cite{barry1993bayesian} Bayesian change point model for this analysis. This model generally assumes different blocks of HR data within the time series of HR in a way that within each block, the mean is constant. The model then calculates the probability of entering a new block as a change point probability. In order to perform BCP on the HR data, we use the \textit{bcp} package written in R programming language \cite{erdman2007bcp}. The BCP is applied to each participant's HR data, and the probability of change at each point is extracted.

After detecting both change points in HR and the presence of certain road objects, we use a window of 10 seconds around each change point in HR to search for the presence of each road object based on the computer vision detection. In the case of the presence of certain road objects, we will search in the next 10 seconds after the detection of the change point for the maximum HR value. Using the HR max values, we define the reaction to the road object as the difference between the HR at the moment of change point and the HR max value. We use Linear Mixed Effect (LME) models to understand the variation in drivers' HR responses around each environmental attribute \cite{brown2021introduction,fox2002linear}. LME models are similar to a simple linear regression with taking into account of the variability across participants in their responses as random factors while accounting for the effect of fixed factors (each perturbation). In the case of facing a count type dependent variable (e.g., number of change points), we use a generalized linear model with a negative binomial process distribution \cite{bolker2009generalized}. The analysis above is performed through the LME4 \cite{bates2007lme4} package written in R programming language \cite{R_lang}.

\subsection{Detection of Drivers' Facial Emotions}\label{sec:affectiva}
In order to detect drivers' facial emotions, we leverage the Affectiva module on the iMotion software \cite{mcduff2013affectiva,mcduff2016affdex}. Previous research has shown the utility of this software in detecting facial expressions and their positivity/negativity level, as well as detecting basic emotions and specific facial muscles \cite{kulke2020comparison,abdic2016driver,mehta2021self,reinares2019cognitive,tavakoli2019multimodal}. For this paper, we focus on the two measures of ``engagement'' and ``valence''. Engagement refers to the level of showing any signs of emotion in the face with a value of 0 (no emotion) to 100 (highest showing of emotion), and valence is a measure of positive or negativity of emotion with a value between -100 (most negative) to +100 (most positive). After performing the analysis with Affectiva, all the frames that did not have any detection were removed from the database. This can be due to the angle of the camera as well as lighting issues, which did not account for a significant portion of the data.       

\subsection{Pixelwise Distance to the Lead Vehicle}\label{sec:distance}
In order to calculate the pixel-wise distance to the lead vehicle, we have used a combination of lane detection and object detection algorithms. For this task, we used a version of the YOLO algorithm titled, YOLOP: You Only Look Once for Panoptic Driving Perception \cite{wu2021yolop}. We applied YOLOP on the outside videos and retrieved lanes, vehicles, and driveable areas. We initially used both lanes and cars, but upon finding YOLOP’s car detection to be inaccurate on cars directly in front in our specific videos, we supplemented the detection with external bounding boxes from MASK RCNN \cite{he2017mask}. Using YOLOP we detect the lane a detected car is located in. The modified program assigns lane numbers to every car detection. Lane 0 represents a car directly in front in the same lane, a negative lane number represents cars to the left, and positive lane numbers represent cars to the right.

\begin{figure*}
  \centering
  \frame{\includegraphics[width=\linewidth]{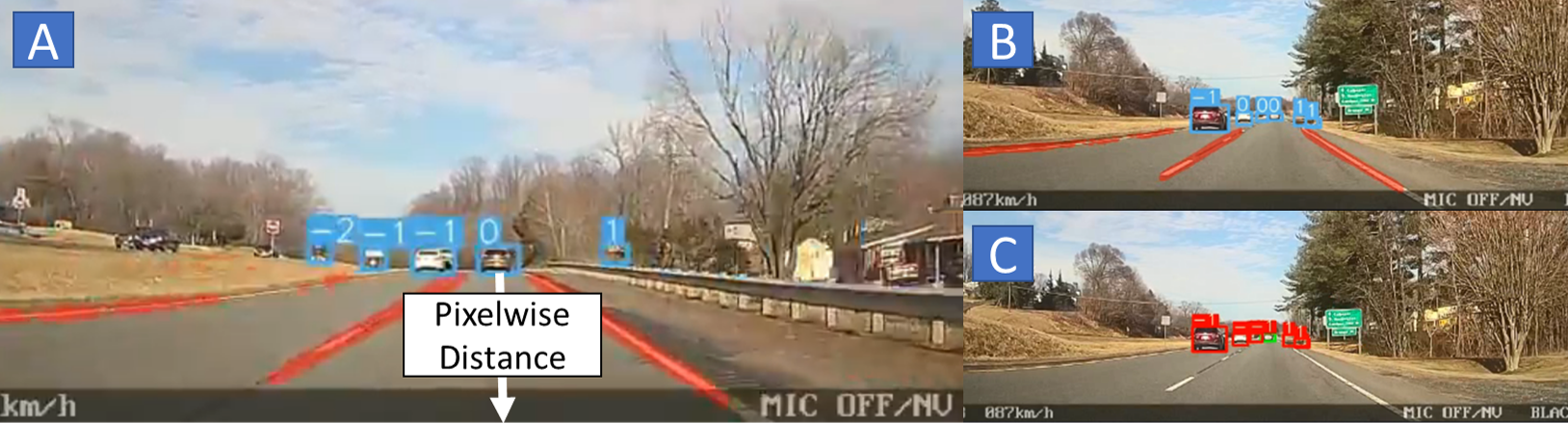}}
  \caption{The methodology to calculate the pixelwise distance from the lead vehicle. (A) detection the lead vehicle using a combination of lane tracking and object detection. Enhancing the high number of false positives in detection as the lead vehicle (B) through post-processing (C)  }
  \label{fig:dist_method}
\end{figure*}

YOLOP’s lane detection creates a pixel mask representing lanes. From the pixel mask, a center is marked, and the amount and location of lanes on either side of the center are found. Then on each car detection, the lane number is assigned to the detection according to the car’s relative position to the lane detections. Additionally, a line is drawn from the bottom center of the mask to the center of each car, and the number of times this line intersects a lane is counted. Zero intersections mean the car will be in the center lane, while one intersection would mean the car is in the line directly adjacent to the current lane (figure \ref{fig:dist_method} - A).

After detection is performed, post-processing is applied to remove outliers and fill in detection in gaps, as car and lane detection will not always provide accurate results, especially when they examine one frame at a time (figure \ref{fig:dist_method} - B). Post-processing also ensures only a single car can have lane number 0 (in front), as only a single car will be visible in front and in the same lane. In some cases, lane detection will not detect lanes at all, and declare all cars visible as the car in front. Post-processing will correct this by assigning the car closest to the center and with the largest bounding box as the car in front while assigning other cars to the side lanes (figure \ref{fig:dist_method} - C).

\section{Results}
\subsection{Relationship Between Road Objects and Drivers' HR and Facial Expressions}
\begin{figure}
  \centering
  \frame{\includegraphics[width=\linewidth]{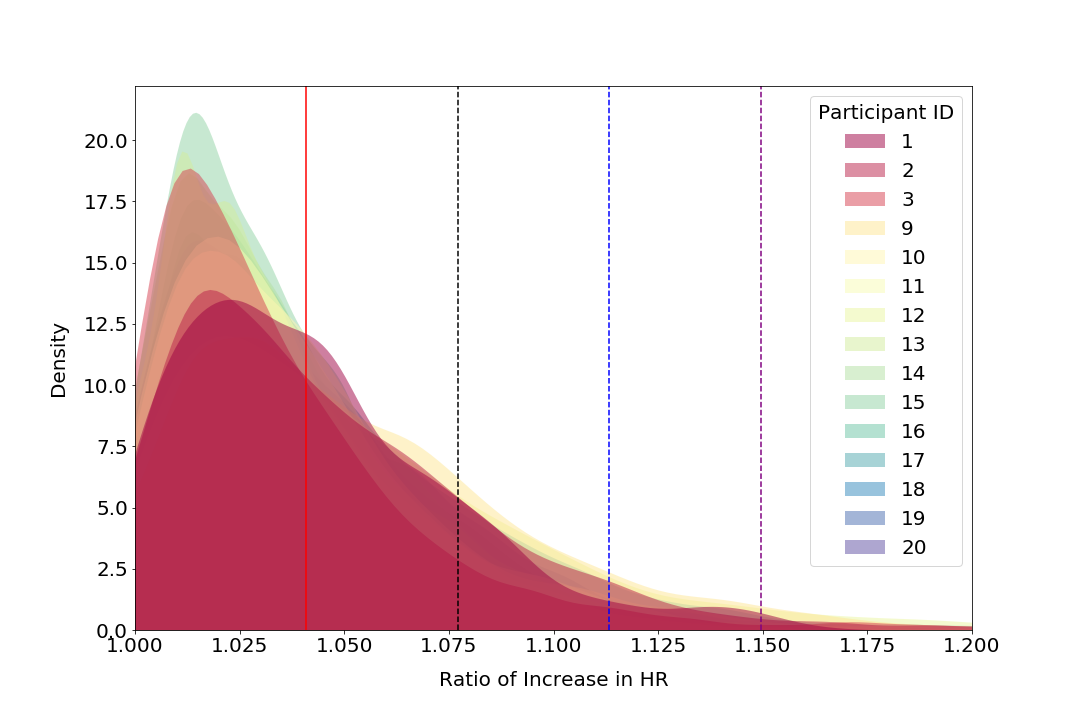}}
  \caption{The distribution of increase in HR at location of change point in percentage for different participants. Note that the red vertical line show the mean, the black dashed line shows the mean + standard deviation, the blue dashed line shows the mean + two * standard deviation, and the purple dashed line shows the mean plus three * standard deviation. }
  \label{fig:hr_distribution}
\end{figure}

Figure \ref{fig:hr_distribution} shows the distribution of the percentage of increase in drivers' HR after each detected change point. We have also marked the location of the mean, as well as one standard deviation, two standard deviations, and three standard deviations by red, black, blue, and purple lines, respectively. Using the values in figure \ref{fig:hr_distribution}, we define different levels of stress to be low, medium, and high based on the level of increase in HR. More specifically, between $\mu$ and $\mu + \sigma$ represents ``low stress'', between $\mu + \sigma$ and $\mu + 2*\sigma$ represents ``medium stress'', and between $\mu + 2*\sigma$ and $\mu + 3*\sigma$ or more represents ``high stress'' level.

Additionally, note that due to the imbalanced nature of the dataset (e.g., unequal number of instances with trucks versus pedestrians), we have performed oversampling based on Synthetic Minority Oversampling Technique (SMOTE) \cite{chawla2002smote} to generate new samples for the minority classes. SMOTE generates new data points from convex combinations of nearest neighbors. This will help us better compare the different categories. Additionally, we have grouped different categories of environmental perturbations for a better illustration. The ``rider'' group contains cyclists and motorcycle riders, and the ``bigger vehicle'' contains trucks and buses. 

\begin{figure}[ht]
  \centering
  \frame{\includegraphics[width=\linewidth]{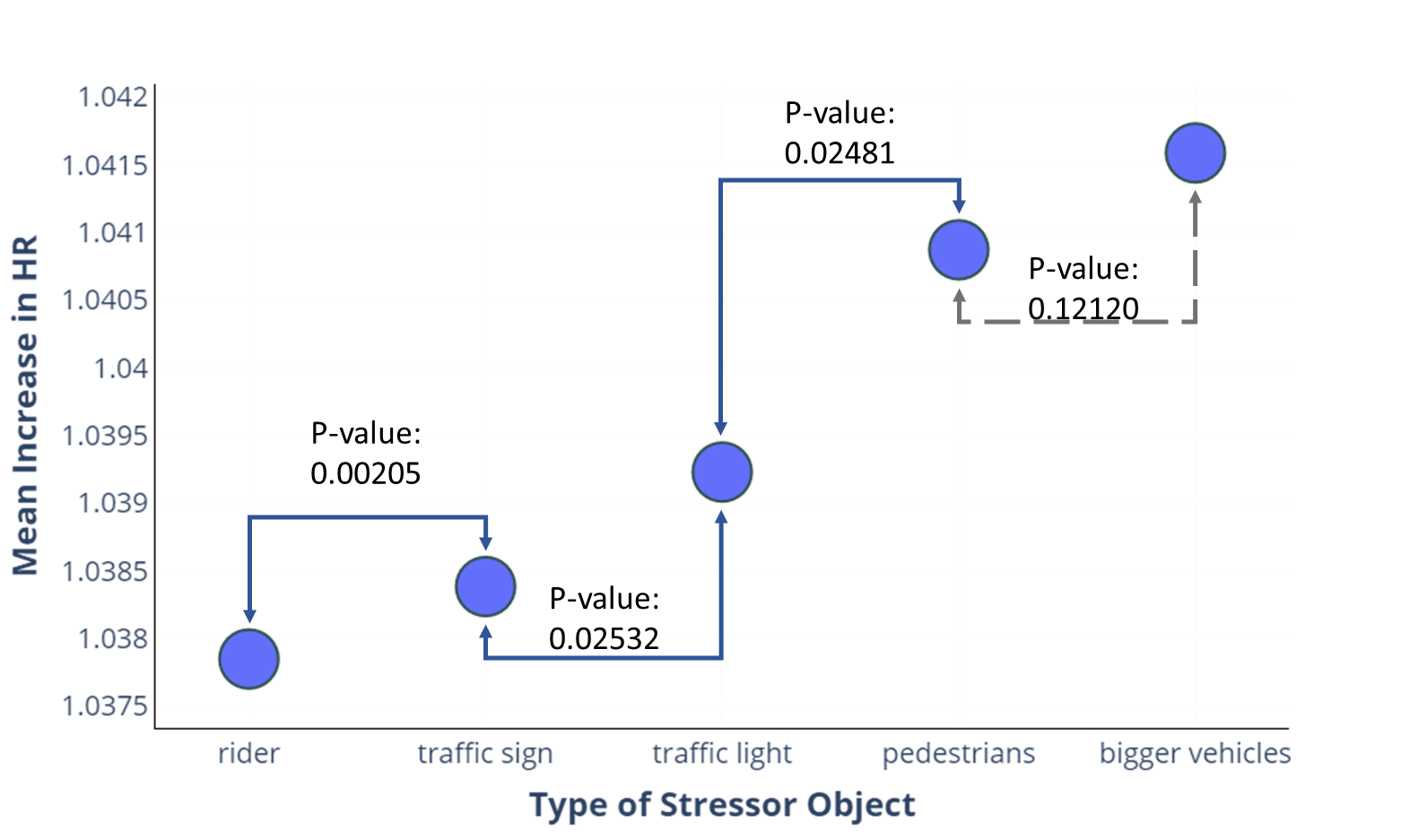}}
  \caption{The average increase in HR at each change point location associated with each road object category.}
  \label{fig:pert_avg}
\end{figure}

Figure \ref{fig:pert_avg} shows the average level of increase in HR for each road object category. The category of the bigger vehicle has the highest amount of increase which includes trucks and buses. This is then followed by pedestrians, traffic signals (an indicator of intersections), traffic signs, and riders. In order better understand the differences across these groups, we run a Kruskal Wallis test \cite{mckight2010kruskal} over the different categories of road stress inducing objects that were associated with HR increase. A Kruskal Wallis test is a non-parametric test that assesses the differences across independent samples. This test shows that the categories of stress-inducing objects are significantly different from each other, with a degree of freedom of 4, a chi-squared value of 34.14, and a p-value of 6.97e-7. We then ran a set of pairwise t-tests that were corrected using the Holm method \cite{holm1979simple}. This is performed due to the fact that multiple comparisons are being made simultaneously. The results of the t-tests are shown in Table \ref{tab:comparison-multiple}. As can be seen, most of the comparisons produce significant results other than the comparison between bigger vehicles and pedestrians categories.      

\begin{table}[ht]
\caption{The comparison between the HR increase associated with different categories of perturbations}
\label{tab:comparison-multiple}
\resizebox{0.48\textwidth}{!}{%
\begin{tabular}{l|llll}
              & Bigger Vehicles  & Pedestrians      & Rider   & Traffic Light \\ \hline
Pedestrians   & 0.12120          & -                & -       & -             \\
Rider         & \textless{}2e-16 & \textless{}2e-16 & -       & -             \\
Traffic Light & 0.00014          & 0.02481          & 1.7e-8  & -             \\
Traffic Sign  & 1.9e-10          & 1.7e-6           & 0.00205 & 0.02532      
\end{tabular}%
}
\end{table}



\begin{figure}[ht]
  \centering
  \frame{\includegraphics[width=\linewidth]{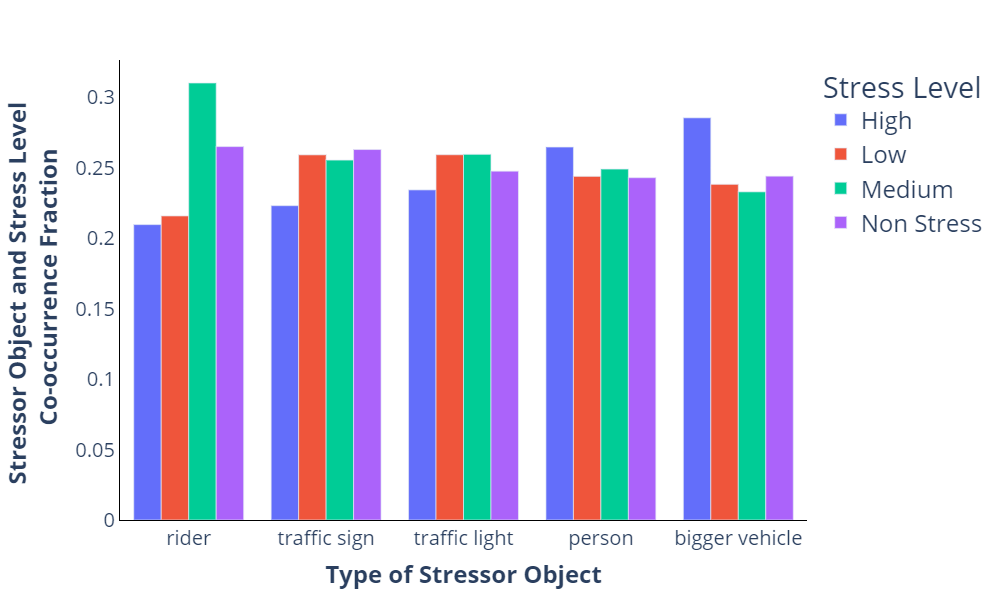}}
  \caption{The fraction of presence of each stress category within each road object.}
  \label{fig:stress_category}
\end{figure}

Figure \ref{fig:stress_category} shows the fraction of the presence of each road object within each stress category. On average, within each stressing object category, the fraction of the presence of each stress level varies, with the category of riders having the lowest fraction of high-stress category and bus and truck (bigger vehicles) having the highest. Although the rider category has the least high-stress level, it leads the medium stress level category. In other words, the presence of a rider is most likely to increase the HR only as much as two standard deviations away from an average increase in HR. Additionally, note that almost 30\% of all the increases in HR associated with trucks move at least two standard deviations away from the average increase. 

Our data shows that there exists differences across participants' HR increase when facing different perturbations. We observe that within different categories of stress-inducing objects, participants objective HR measures are different from each other. For example, participant \#14 has the least increase in HR due to bigger vehicles, while participant \#19 has the highest level.  

\begin{figure}
  \centering
  \frame{\includegraphics[width=\linewidth]{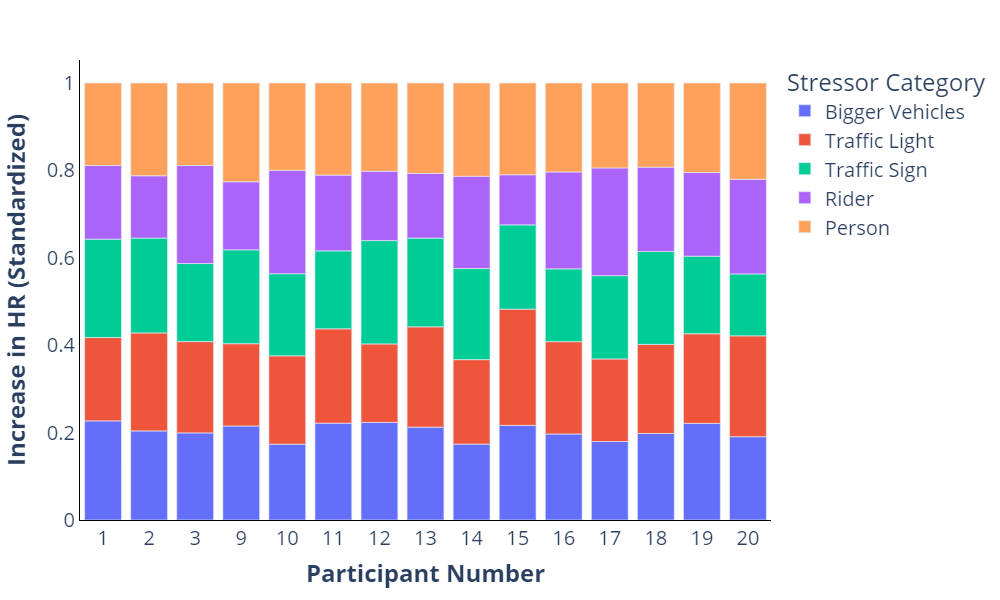}}
  \caption{A comparison between the differences across participants for various stress-inducing object categories on the road.}
  \label{fig:comp_all_par}
\end{figure} 



For valence and engagement, we have followed a similar analysis procedure as HR. We have first assigned categories to valence and engagement based on the mean and standard deviation of valence and engagement. Valence is categorized as negative, neutral, and positive, where the range between $\mu \pm \sigma$ is considered neutral. This is mostly due to the fact that valence values are often close to zero (showing no emotion). For engagement, we considered two categories of neutral and non-neutral engagement. All the values more than $\mu + \sigma$ are considered as non-neutral facial engagement. 

\begin{figure}[ht]
  \centering
  \frame{\includegraphics[width=\linewidth]{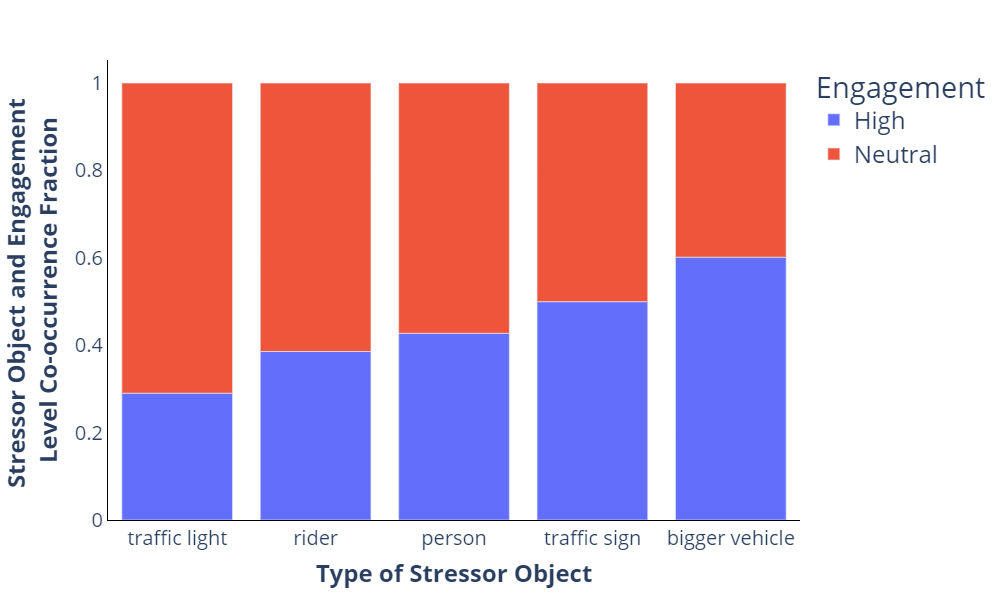}}
  \caption{The presence of each road object in each category of engagement. Note that we provide a zoomed in version for the categories with lower values.}
  \label{fig:eng_category}
\end{figure}

Figure \ref{fig:eng_category} shows the presence of each stress inducing object category within each engagement level category. Note that similar to HR increases, bigger vehicles have the highest amount of high engagement, and rider is among the lowest categories. It is interesting that the proportion of having high facial engagement is more than 0.5 for the bigger vehicle category, which implies that when facing bigger vehicles, drivers are more likely to show some level of facial expression, while this value is the lowest for the traffic signal category. In other words, detecting responses to traffic signal might be more feasible by using HR rather than facial expressions.

\begin{figure}[ht]
  \centering
  \frame{\includegraphics[width=\linewidth]{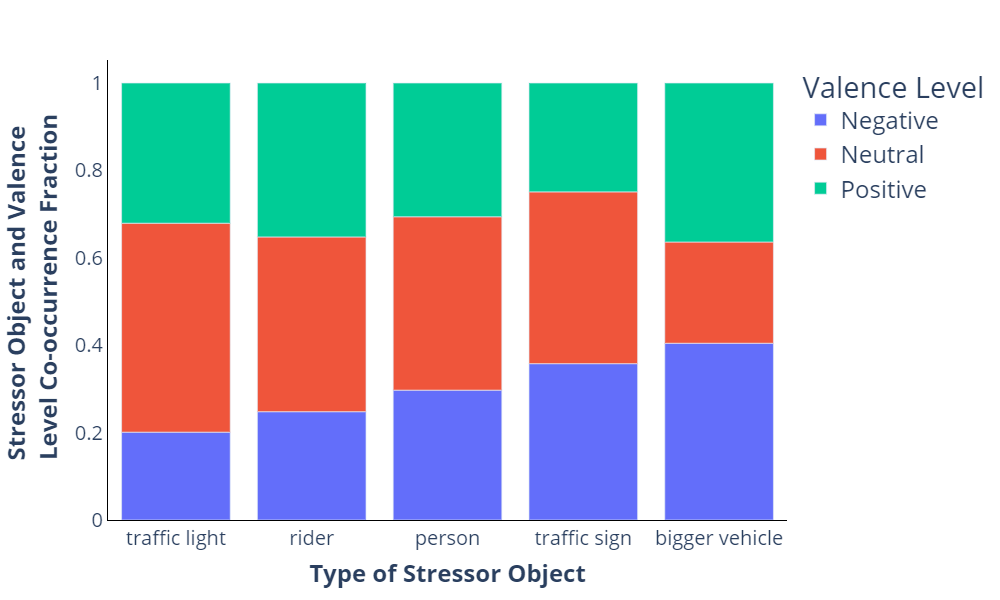}}
  \caption{The presence of each road object in each category of valence. Note that we provide a zoomed in version for the categories with lower values.}
  \label{fig:val_category}
\end{figure}

Figure \ref{fig:val_category} shows the presence of each stress inducing category within each valence level category. Similar to both engagement and HR changes, bigger vehicles are among the categories with the highest level of negative valence. Similar to facial engagement, the traffic signal is mostly followed by a neutral facial expression which might indicate that this modality may not be suitable for detecting reactions to the traffic signals. 

\subsection{A Detailed Analysis of HR in the Vicinity of Traffic Signs}

In addition to the generic traffic sign detection, we have analyzed drivers' HR in the vicinity of stop signs and speed limit signs from the pool of detected traffic signs. These two traffic signs are categorized as regulatory signs, which might have a different effect on the driver as compared to the other signs. In this section, we are mostly interested in knowing the physiological pattern of drivers around these two traffic signs. In order to find clusters in drivers' physiological metrics, we perform k-means clustering on the HR signal around these two regulatory signs.  

It is important to note that changes in human HR and, in general physiology can happen at different time scales with respect to the detection of certain road objects. For example, we don't exactly know when (e.g., how many seconds prior or after presence or detection of an object) a person might perceive a road object and react. Thus in order to compare the time series and further cluster them, we need to first bring different instances of the stop and speed limit sign occurrences into the same time frame using Dynamic Time Warping (DTW).

DTW is a technique for finding an optimal warping function to transform a time series to another one that might have differences in speed of happening in time \cite{muller2007dynamic,berndt1994using}. For example, imagine the HR of participants when reaching a stop sign. Different participants might reach and pass through a stop sign with different duration, thus producing time series with different durations for the same event. By applying DTW, we create time series of approaching stop signs or speed limit signs with similar durations. We perform DTW using the \textit{tslearn} package programmed in Python \cite{JMLR:v21:20-091}.

After performing DTW on the time series of HR in the vicinity of stop and speed limit signs, we performed k-means clustering. K-means clustering is an unsupervised approach to finding clusters within the data \cite{jain1999data}. This algorithm which lies under the partition-based clustering methods, performs based on assigning each point to a randomly initialized set of partitions based on their similarity. This procedure is performed until convergence. 

In order to define the number of clusters needed, we use the silhouette score \cite{shahapure2020cluster}. This score shows the quality of clustering by measuring how close points from different clusters are to each other \cite{shahapure2020cluster}. Silhouette score is a value between -1 and 1, which 1 indicates the most separation between clusters. We assess the Silhouette score for the different number of clusters and choose the number of clusters that produce the highest score, which in both cases of being close to a stop sign or speed limit sign, two clusters have been chosen.

\begin{figure}[ht]
  \centering
  \frame{\includegraphics[width=\linewidth]{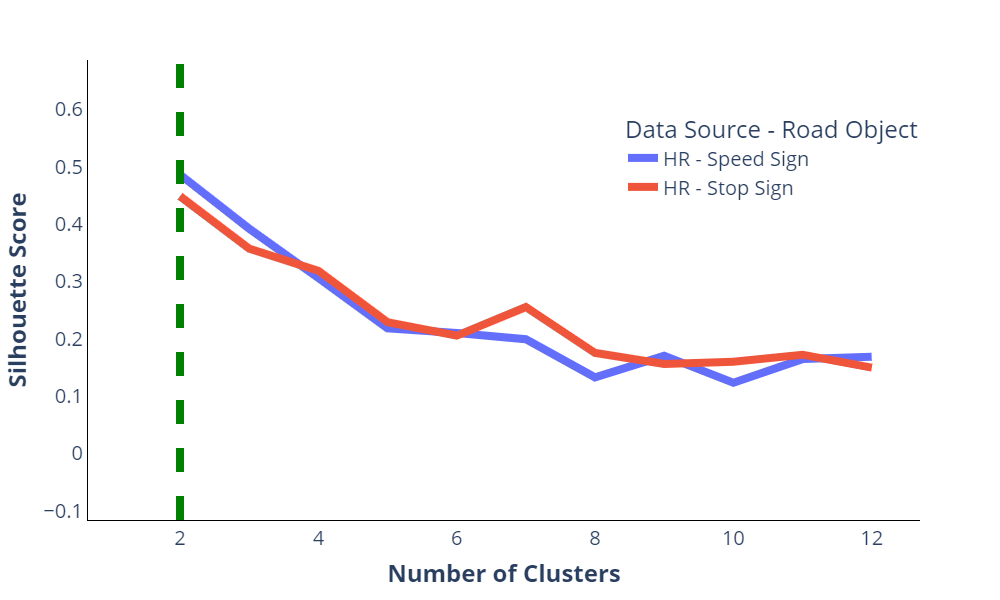}}
  \caption{The Silhouette score for different number clusters for k-means clustering of drivers' HR in the vicinity of speed limit signs and stop signs. The optimum cluster number is 2} 
  \label{fig:s-speed}
\end{figure}

\begin{figure}[ht]
  \centering
  \frame{\includegraphics[width=\linewidth]{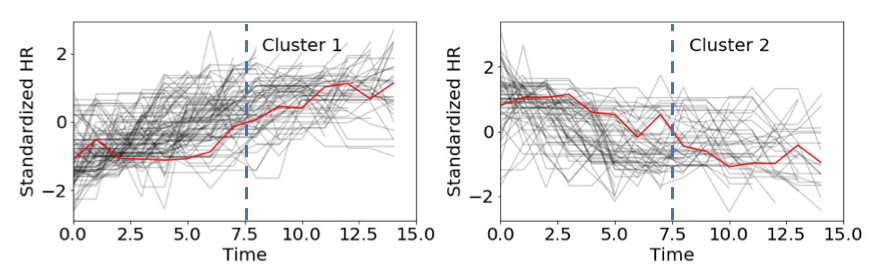}}
  \caption{The patterns in drivers' HR in the vicinity of speed limit signs. The two clusters both have an abrupt increase in HR in the vicinity of speed limit sign. The dashed blue line shows the moment of speed limit sign detection} 
  \label{fig:speed_hr_clust}
\end{figure}

Based on the two clusters detected as the optimal number of clusters, we apply the k-means clustering. Figure \ref{fig:speed_hr_clust} shows the two clusters detected. Note that x-axis in the middle is the time of reaching a speed limit sign based on computer vision detection (Time = 7.5). While the two patterns around the speed limit signs are very different from each other at the first look, they have certain characteristics. Cluster 1 is related to the cases that the HR is at its normal value prior to reaching the speed limit sign. Note the abrupt increase in HR in the vicinity of the sign. This cluster happens more frequently compared to the other cluster. Cluster 2 is related to the cases where the HR is already higher than drivers' normal (baseline) prior to arriving at the speed limit sign. It is interesting that even in this case, a small increase is observed at the location of speed limit sign detection (time =7.5).

\begin{figure}[ht]
  \centering
  \frame{\includegraphics[width=\linewidth]{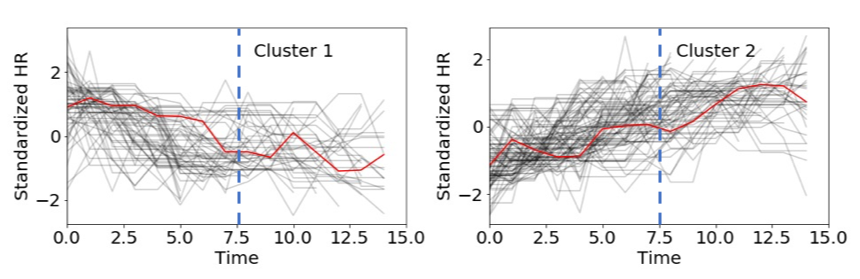}}
  \caption{The patterns in drivers' HR in the vicinity of speed limit signs. The two clusters both have an abrupt increase in HR in the vicinity of stop sign. The dashed blue line shows the moment of stop limit sign detection} 
  \label{fig:stop_hr_clust}
\end{figure}

The patterns in the vicinity of stop signs have a similar trend to the speed limit signs with some interesting differences (figure \ref{fig:stop_hr_clust}). First, in cluster 1, the increase in HR happens a few seconds after the detection of stop signs, which can mean a time difference between the effect of these two signs on human physiology. Second, the stop sign trend in cluster 2 has a downward trend after time =12.5 (see figure \ref{fig:stop_hr_clust} , time $>$ 12.5), whereas this is not the case for cluster 1 in speed limit (figure \ref{fig:speed_hr_clust} - time $>$ 12.5). This might indicate that the speed limit sign has a more prolonging effect than the stop sign on human physiology. 

\subsection{Relationship Between Distance to the Lead Vehicle and HR}
\begin{figure}[ht]
  \centering
  \frame{\includegraphics[width=\linewidth]{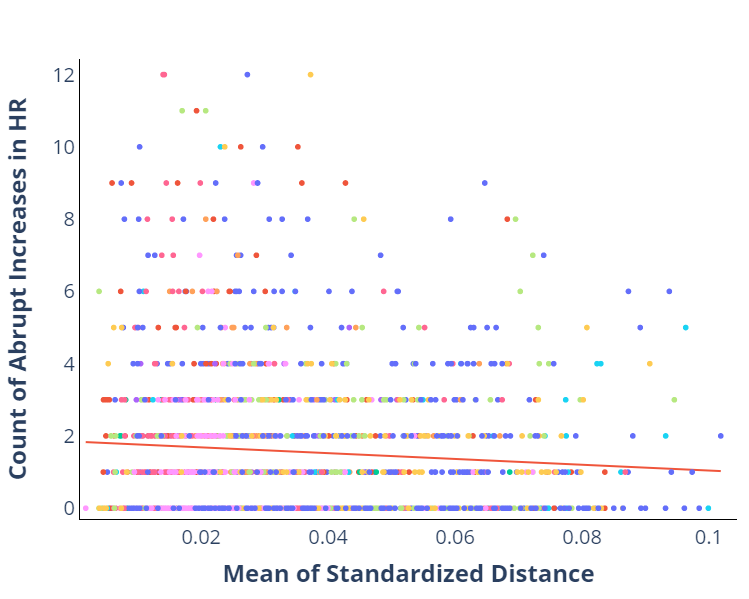}}
  \caption{The change in the average distance to the lead vehicle versus the number of abrupt increases in drivers' heart rate. The markers show the datapoints for each participant, and the line plot is the overall trendline.}
  \label{fig:hr_mean}
\end{figure}

Figure \ref{fig:hr_mean} shows the number of abrupt increases in drivers' HR versus the average distance to the lead vehicle. Visual inspection of the figure suggests that a negative relationship exists between the average distance to the lead vehicle and the count of abrupt increases in drivers' HR. We test this relationship by using a Generalized Linear Mixed-effect Model (GLMM) with a negative binomial process, in which the independent variable is the average distance to the lead vehicle and the dependent variable is the count of the number of abrupt increases in HR. This model considers the random effect of different participants' baselines with separate intercepts for each participant. We chose this model specifically by comparing it with a generalized model with no random effect. We chose the aforementioned model based on the Akaike's information criterion (AIC) model comparison \cite{wagenmakers2004aic}. As shown on Table \ref{tab:model-comp-mean}, the GLMM model has lowest AIC value.

\begin{table}[]
\caption{The comparison between the two models chosen for modeling the relationship between abrupt increases in HR and average distance to lead vehicle. The lowest AIC in GLMM with random intercept is the basis for model selection.}
\label{tab:model-comp-mean}
\resizebox{0.48\textwidth}{!}{%
\begin{tabular}{lllllll}
Model Name                 & AIC    & BIC    & Loglikelihood & Chi sq & Df & Pr       \\ \hline
Generalized Linear Model   & 6256.8 & 6273.3 & -3125.4       &        &    &          \\
GLMM with random intercept & 6246.4 & 6268.4 & -3119.2       & 12.397 & 1  & 0.00043
\end{tabular}%
}
\end{table}

Table \ref{tab:distance-model_mean} shows the result of the chosen GLMM model for the mean distance to lead vehicle data. The results show that participants' HR had a higher number of abrupt changes as the mean distance to the lead vehicle decreased. 

\begin{table}[]
\caption{The result of applying the GLMM model with random slope on the standard deviation of distance to lead vehicle versus abrupt increases in drivers' HR. p-values estimated via t-tests using the Satterthwaite approximations to degrees of freedom. The significant predictors at 0.05 level are underlined in the Pr column.}
\label{tab:distance-model_mean}
\resizebox{0.48\textwidth}{!}{%
\begin{tabular}{lllllll}
Effect    & Estimate & Std. Error & z-value     & Pr(\textgreater{}{[}z{]}) & CI 2.5\%  & CI 97.5\% \\ \hline
Intercept & 0.38707      & 0.06511       & 5.945 & {\ul{2.7e-9}}                  & 0.23752773    & 0.5160669  \\
Distance  & 0.07928  & 0.03572    & 2.220 & {\ul{0.0264}}                  & 0.01033641 & 0.1504898
\end{tabular}%
}
\end{table}

\begin{figure}[ht]
  \centering
  \frame{\includegraphics[width=\linewidth]{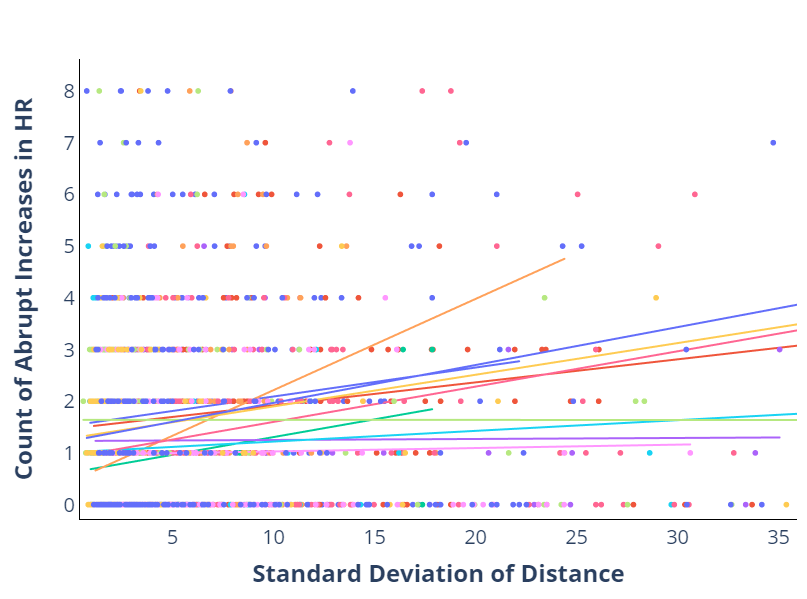}}
  \caption{The change in the standard deviation of distance to the lead vehicle versus the number of abrupt increases in drivers' heart rate. The markers show the datapoints for each participant, and the line plots are the best fitted line for each participant's data.}
  \label{fig:hr_dist}
\end{figure}

Figure \ref{fig:hr_dist} shows the number of abrupt increases in drivers' HR versus the standard deviation of distance to the lead vehicle. Visual inspection of the figure suggests that a positive relationship exists between the standard deviation in the distance to the lead vehicle and the count of abrupt increases in drivers' HR. We test this relationship by using a generalized linear mixed-effect model (GLMM) with a negative binomial process, in which the independent variable is the standard deviation of the distance to the lead vehicle. The dependent variable is the count of the number of abrupt increases in HR while considering the random effect of different participants' baselines with separate intercepts for each participant. We chose this model specifically by comparing it with a generalized model with no random effect. We chose the aforementioned model based on the AIC model comparison. As shown in Table \ref{tab:model-comp}, the GLMM model has the lowest AIC value.   

\begin{table}[]
\caption{The comparison between the two models chosen for modeling the relationship between abrupt increases in HR and standard deviation of distance to lead vehicle. The lowest AIC in GLMM with random intercept is the basis for model selection.}
\label{tab:model-comp}
\resizebox{0.48\textwidth}{!}{%
\begin{tabular}{lllllll}
Model Name                 & AIC    & BIC    & Loglikelihood & Chi sq & Df & Pr       \\ \hline
Generalized Linear Model   & 7335.1 & 7346.1 & -3665.5       &        &    &          \\
GLMM with random intercept & 7272.5 & 7289.0 & -3633.2       & 64.590 & 1  & 9.22e-16
\end{tabular}%
}
\end{table}

Table \ref{tab:distance-model} shows the result of the chosen GLMM model for the distance to lead vehicle data. The results show that participants' HR had a higher number of abrupt changes as the distance to the lead vehicle changed more sporadically (standard error = 0.01235, z-value = 11.308, p-value $<$ 0.0001). 

\begin{table}[]
\caption{The result of applying the GLMM model with random slope on the standard deviation of distance to lead vehicle versus abrupt increases in drivers' HR. p-values estimated via t-tests using the Satterthwaite approximations to degrees of freedom. The significant predictors at 0.05 level are underlined in the Pr column.}
\label{tab:distance-model}
\resizebox{0.48\textwidth}{!}{%
\begin{tabular}{lllllll}
Effect    & Estimate & Std. Error & z-value     & Pr(\textgreater{}{[}z{]}) & CI 2.5\%  & CI 97.5\% \\ \hline
Intercept & 0.43270      & 0.01895       & 22.83 & {\ul{2e16}}                  & 0.21162163    & 0.4832381  \\
Distance  & 0.16748  & 0.01327    & 12.62 & {\ul{2e-16}}                  & 0.08054035 & 0.3128414
\end{tabular}%
}
\end{table}

\subsection{Relationship between Drivers' Speed and HR and Facial Expressions}

\begin{figure}
  \centering
  \frame{\includegraphics[width=\linewidth]{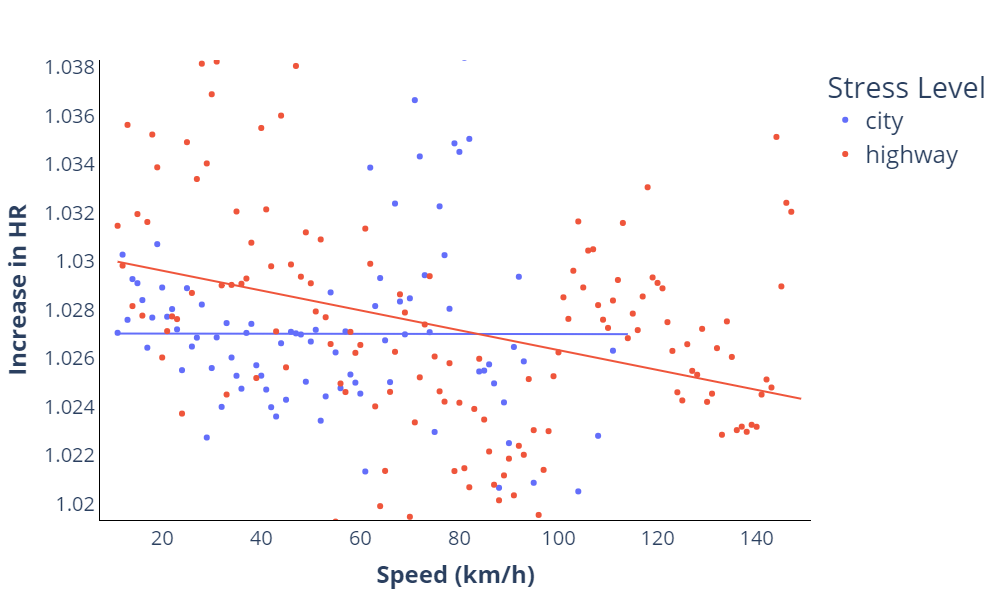}}
  \caption{The change in heart rate at the locations of changepoints with respect to varying levels of speed for city and highway environments.}
  \label{fig:speed_bcp}
\end{figure}

Lastly, we analyze the relationship between the drivers' speed and the rate of increase in HR as well as changes in facial expressions. Figure \ref{fig:speed_bcp} shows the change in HR in the vicinity of change points in HR for two different environments of the city versus highway. While both environments show a declining trend with respect to the relationship between HR and speed (as speed increases, the rate of increase in HR decreases), a higher slope is observed for the highway environment. In this section, we treated the environment type as a random factor and assessed the relationship with a mixed effect model. Table \ref{tab:speed-model} shows the result of the mixed effect model. Also, note that we chose this model as it had a lower AIC value compared to a simple linear regression (-256158 versus -256140).

\begin{table}[]
\caption{The result of the linear mixed effect model on the changes in HR at different speed in city and environment. p-values estimated via t-tests using the Satterthwaite approximations to degrees of freedom.}
\label{tab:speed-model}
\resizebox{0.48\textwidth}{!}{%
\begin{tabular}{llllll}
Effect    & Estimate & Std. Error & df     & t-value & Pr   \\ \hline
Intercept & 1.042  & 1.736e-3    & 1.041e+00 & 600.086                  & {\ul{0.000825}}    \\
Speed  & -2.839e-05  & 4.688e-06    & 1.030e+04 & -6.056                  & {\ul{1.44e-09}} 
\end{tabular}%
}
\end{table}

\begin{figure}
  \centering
  \frame{\includegraphics[width=\linewidth]{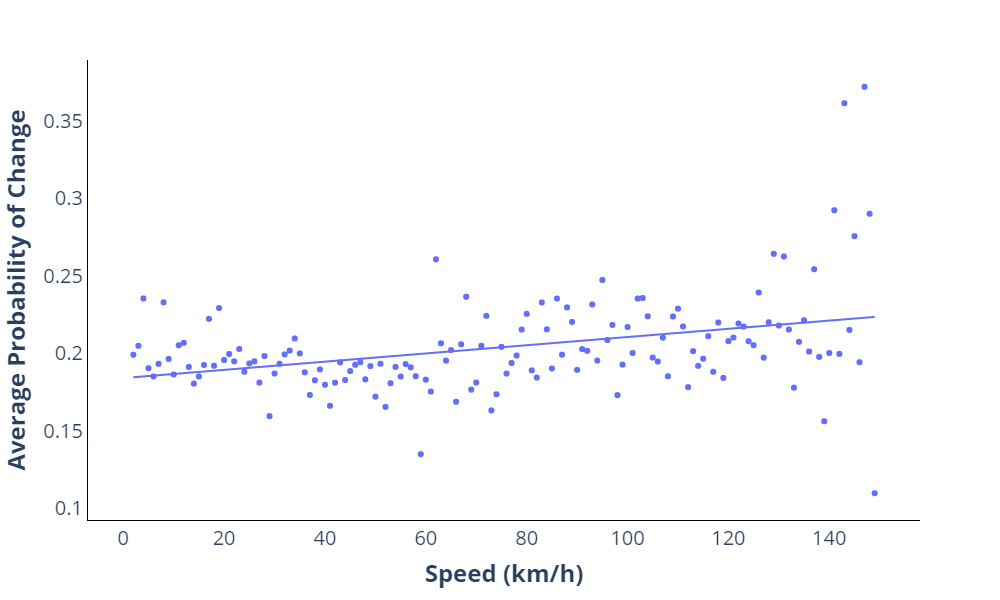}}
  \caption{The probability of increasing change in HR at different speeds}
  \label{fig:speed_hr_bcp}
\end{figure}

Figure \ref{fig:speed_hr_bcp} shows the probability of change in HR at different speeds of the vehicle. As it is shown, higher speeds are accompanied by higher probabilities of abrupt changes in drivers' HR towards higher values. We tested this relationship using a linear regression model, and the results are shown in Table \ref{tab:speed-model-bcp}.   

\begin{table}[]
\caption{The result of the linear regression model on evaluating changes in probability of abrupt increases in HR based on drivers' speed}
\label{tab:speed-model-bcp}
\resizebox{0.48\textwidth}{!}{%
\begin{tabular}{lllllll}
Effect    & Estimate & Std. Error & t-value     & Pr & CI 2.5\%  & CI 97.5\% \\ \hline
Intercept &1.839e-01      & 4.963e-03       & 37.058 & {\ul{2e-16}}                  & 0.1741    & 0.1937  \\
Speed  & 2.651e-04  & 5.721e-05     & 4.633 & {\ul{7.91e-06}}                  & 0.0001 &  0.0004
\end{tabular}%
}
\end{table}

\begin{figure}
  \centering
  \frame{\includegraphics[width=\linewidth]{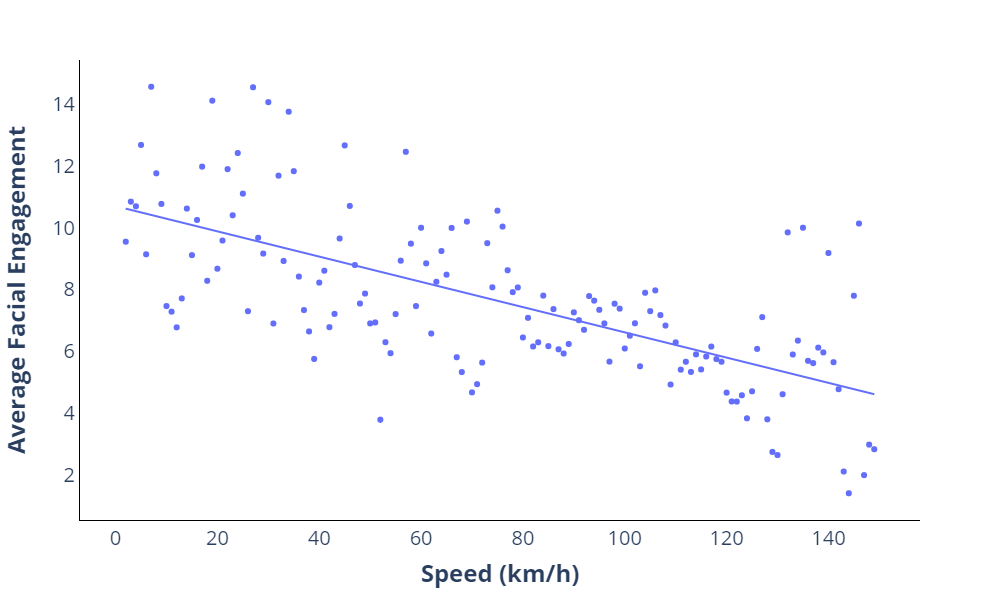}}
  \caption{The facial engagement with respect to varying levels of speed}
  \label{fig:speed_eng}
\end{figure}

Figure \ref{fig:speed_eng} shows the relationship between drivers' facial expressions and speed. On average, drivers are less likely to show a specific movement in their facial muscles as they travel at a higher speed. The result of testing this relationship with linear regression is also shown in Table \ref{tab:speed-model-eng}

\begin{table}[]
\caption{The result of the linear regression model on evaluating changes in drivers' facial engagement based on drivers' speed}
\label{tab:speed-model-eng}
\resizebox{0.48\textwidth}{!}{%
\begin{tabular}{lllllll}
Effect    & Estimate & Std. Error & t-value     & Pr & CI 2.5\%  & CI 97.5\% \\ \hline
Intercept & 10.694699       & 0.328603       &  32.55 & {\ul{2e-16}}                  & 10.0452    & 11.3441  \\
Speed  & -0.040836  & 0.003788     & -10.78 & {\ul{2e-16}}                  & -0.0483 &  -0.0333
\end{tabular}%
}
\end{table}

\begin{figure}
  \centering
  \frame{\includegraphics[width=\linewidth]{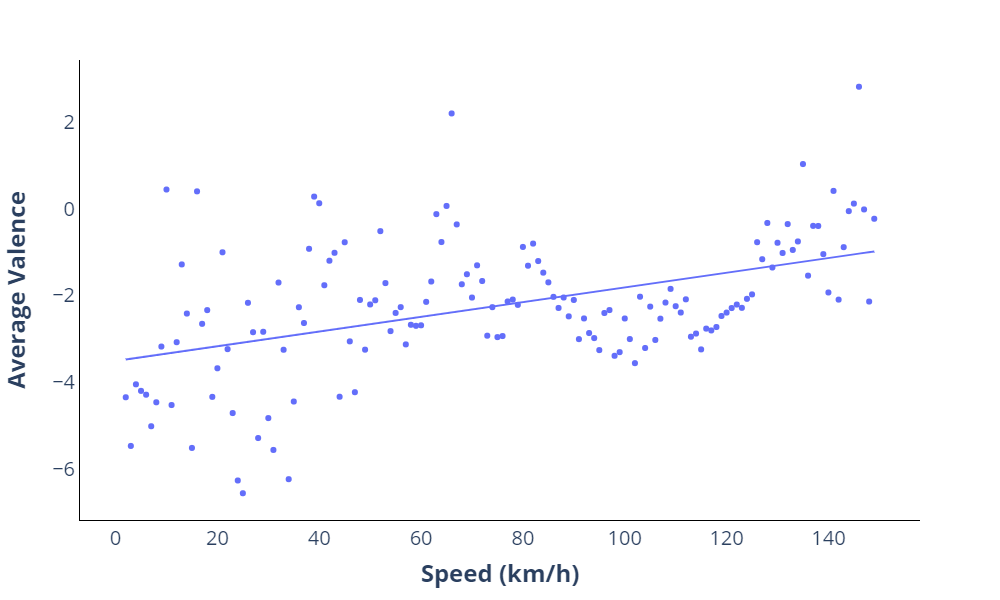}}
  \caption{The facial valence with respect to varying levels of speed }
  \label{fig:speed_val}
\end{figure}

Figure \ref{fig:speed_val} shows the relationship between drivers' valence at varying levels of speed. As shown, drivers' valence has a more positive value at higher speeds. In other words, drivers show more positive facial expressions when they drive at higher speeds on highways. This result is also tested by using a linear regression model as shown in Table \ref{tab:speed-model-val}.

\begin{table}[]
\caption{The result of the linear regression model on evaluating the relationship between in drivers' valence on drivers' speed}
\label{tab:speed-model-val}
\resizebox{0.48\textwidth}{!}{%
\begin{tabular}{lllllll}
Effect    & Estimate & Std. Error & t-value     & Pr & CI 2.5\%  & CI 97.5\% \\ \hline
Intercept &  0.016972       & 0.002657       &  6.388 & {\ul{2.11e-9}}                  & -3.9709    & -3.0600  \\
Speed  & -0.040836  & 0.003788     & -10.78 & {\ul{2e-16}}                  & 0.0117 &  0.0222
\end{tabular}%
}
\end{table}

\section{Discussion}
In this study, we undertook an exploratory approach to the relationship between drivers' HR, facial expressions, and driving context, such as the presence of certain road objects and distance to lead vehicles. As a summary, our results suggest significant relationships between changes in drivers' psychophysiological measures in the vicinity of the aforementioned road object categories. Our results, while collected objectively, are in line with previous studies that were performed through controlled experiments and by using subjective measures such as self-reports. 

Previous research showed that certain road object categories were accompanied by very high subjective stress levels. For instance, \cite{bustos2021predicting} showed that riders and big vehicles were accompanied by a high fraction of high stress levels. Additionally, studies such as \cite{dittrich2021drivers} note that intersections are accompanied by very high subjective negative emotions. While focusing on drivers' HR instead of subjective measures, our results show similar patterns. Based on the changes in drivers' HR in the vicinity of such road objects, bigger vehicles and pedestrians are among the highest increase in HR, followed by traffic signals, traffic signs, and riders. Moreover, changes in human HR in the vicinity of these objects might fall into different categories of increases in HR. For example, trucks and, in general, bigger vehicles are associated with a higher proportion of the increase in HR with more than two standard deviations away from the mean (high stress), whereas riders are often associated with a medium increase in HR (not more than two standard deviation increase in HR on figure \ref{fig:stress_category}). These results have strong implication for designing human-centered autonomous systems that may need to reason and decide for choosing routes (e.g., a route with a higher number of trucks such as highways versus a route with higher riders), as well as following or passing a road object (e.g., bicycle versus truck). 

Additionally, we observe similar patterns within the engagement and valence of the drivers, where trucks and buses are among the top categories when comparing the negative emotion proportions, which is followed by traffic signs, pedestrians, riders, and traffic signals. While for some categories of road objects, the result of HR and facial expressions confirm each other (e.g., a high fraction of high-stress HR and negative valence for bigger vehicles), this is not necessarily the case for some of the other categories. For example, while traffic signals and, in general, intersections were previously shown to be associated with higher stress levels, and we observe a higher fraction of HR increases beyond two standard deviations from the mean; they exhibit the lowest fraction of high facial engagement as well as the lowest proportion of negative valence as shown on figures \ref{fig:eng_category}. Results as such might indicate that not always participants show facial expressions when they experience increases in HR. This implies the importance of multi-modal sensing for human emotion and stress detection.   

Analysis of patterns in drivers' HR in the vicinity of traffic signs shows that even within signs, there can be similarities and differences in how drivers react objectively to each sign. For instance, we observe that within each of the two traffic signs, two patterns can be detected where in both of them, drivers' HR has an abrupt increase in the vicinity of them. Additionally, we observe that even similar clusters, when comparing stop signs and speed limit signs, might have a different prolonging effect on drivers' physiology. Similar analysis should be performed for other road objects to better understand the pattern in HR in the vicinity of these objects (e.g., the difference between different types of trucks). 

Our results show that a shorter distance to the lead vehicle is associated with a higher number of abrupt increases in drivers' HR. Additionally, we showed that a higher standard deviation of the distance with the lead vehicle is associated with higher levels of abrupt increases in HR. Previous research showed a positive correlation between increases in HR and stress levels. Taken the above together, our results indicate that being closer to the lead vehicle as well as changing the distance to abrupt values away from the mean may be accompanied by higher stress levels and unhealthy states for the driver. These results are in line with recent studies showing that within a driving simulator, shorter time headways are associated with higher workloads, which might affect drivers' safety \cite{radhakrishnan2022physiological}. The fact that not only the average distance but also the standard deviation of distance has an impact on drivers' HR highlights the importance of connected vehicles where the distance can be calibrated and kept constant based on different user profiles. 

We also observed that different participants might be affected differently by each of the road objects as well as different distances to the lead vehicle. For example, we observe the difference between participants \#12 and \#17 on how they respond to bigger vehicles based on their HR (figure \ref{fig:comp_all_par}). This has implications for designing personalized systems that can respond to each individual based on their specific profile in each specific context. In other words, our results indicate the importance of avoiding one-size-fits-all models and developing personalized models. The current scheme of designing in-cabin systems often ignores the individual differences across people. Our results warrant more human-centered considerations with an individual profile approach for designing in-cabin systems that can understand how each user might be affected by environmental attributes. 

Our research show that higher speed, especially in highway environments, is correlated with lower levels of increase in drivers' HR. Previous research provided evidence on the fact that highways might be associated with lower levels of subjective stress levels \cite{dittrich2021drivers,tavakoli2020personalized}. This might indicate that reason behind perceiving highways as a less stressful environment might lie in the fact that drivers are often allowed to drive faster in the highway environment. Note that the negative correlation between speed and stress level (as measured by an increase in HR) is not as strong in the city environment. This might indicate that although in the city environment, higher speed might contribute to lowering the stress level, the presence of other stress inducing objects (e.g., traffic lights, pedestrians, riders, stop signs, etc) might compensate for the increase in speed. Additionally, similar results were found in the recent work of \cite{milardo2021understanding} where higher speed was more correlated with lower standard deviation in HR. Moreover, we showed that higher speeds are correlated with a higher probability of change in HR. In other words, at a higher speed, there is a higher chance that drivers' HR faces an abrupt increase. Taken the two together, we can conclude that at a higher speed, the probability of change is higher, but such a change only has a minor effect on the HR. 

The importance of different road objects and driving context as a whole is highlighted in keeping a less stressful driving experience. In the context of future vehicles, our results have implications for designing autonomous vehicles that can take actions and decisions that are a better fit for a passenger. For instance, in the case of keeping the optimal distance to the lead vehicle, drivers' HR can be used as an indicator for stress level detection at different distances. In the context of routing, our results have implications for designing human-centered routing systems that can provide options based on users' predicted feelings within each route with respect to different objects and road characteristics that might be present within different roads \cite{tavakoli2021leveraging}.

\section{Limitations}
This study has a number of limitations. First and foremost, the number of participants can be increased. Increasing the poll of participants not only helps with finding individual differences to a greater extent, but it also lays the ground to find differences across age groups, different genders, and also socioeconomic backgrounds in their reaction to each road environment attribute. 

Additionally, we note that this work lies heavily on the off-the-shelf computer vision algorithms and models that were developed on certain datasets that may not necessarily represent the proper set of road stress inducing objects. For example, it can be the case that participants had different reactions to different types of trucks (e.g., trailers versus regular trucks). Part of the future work should focus on developing an object detection model that can extract such features from the massive amount of collected naturalistic videos. 

While previous studies have used automatic facial expression recognition software, it is not a general consensus that such software applications can truly and accurately extract all the facial expressions. This is especially more experienced with pre-recorded in-the-wild videos where the angle of the camera, lighting, and other camera-related characteristics can change as the driving happens. Items as such can degrade the accuracy of the facial expression software and also affect the result. While we leveraged data cleaning methods to remove the unusable frames from the facial expression detection point of view, and manually inspected the results and ensured all of our videos were in daylight condition, future work should investigate the effect of such matters in greater detail.  

In this research we used computer vision to detect the distance to the lead vehicle; however, it is intuitive that not always the pixel wise distance represents the true distance to the vehicle. For example, within different road curves, as well as uphill and downhill driving, the distance can be very different than the visual distance in the camera. Future work should integrate other sensing modalities and a higher level of computer vision integration to detect the distance to lead vehicles with higher accuracy. 

Within different modalities, there can be a lag in how each modality responds to a stimulus. For instance, it can be the case that HR increase happens faster than changes in facial expressions. In one of our previous works, we have seen possible lags between different latent constructs (i.e., cognitive load and stress) \cite{tavakoli2021statespace}. While in this paper we analyzed the different modalities, we note that a limitation is to consider the possible lag between modalities for a more proper comparison. Additionally, the response to different environmental perturbations might also be different from a time scale point of view. It can be the case that there is a delay in response to the presence of a truck in comparison to a traffic signal. Part of the future work should be focused on understanding the lag in responding to each environmental perturbation between different sensing modalities.

\section{Conclusion}
This research takes an exploratory approach to understand the effect of road environment objects on drivers' psychophysiological metrics. In contrast to previous studies, this research aims to understand the reason behind changes in drivers' emotions and feeling, which can later be used for interventional purposes. By analyzing naturalistic driving data from 15 participants within varying city and highway environments, we find that different road objects might be associated with different levels of increase in drivers' HR as well as different proportions of negative facial emotions. Our results indicate that bigger vehicles are associated with the highest amount of increase in drivers' HR as well as negative emotions. Additionally, we showed that shorter distances to the lead vehicle in naturalistic driving as well as the higher standard deviation in the distance might be associated with a higher number of abrupt increases in drivers' HR, showing a possible increase in stress level. Lastly, our results indicate, on average, while driving at a higher speed, participants showed a higher levels of positive emotions, increase in their facial engagement, followed by a lower rate of increase in HR.

\section{Appendix}
\subsection{Linear Mixed Effect (LME) Model}
Based on the notation provided in \cite{fox2002linear}, we can define a LME model as:

\begin{equation}\label{lme}
    y = X\beta + b_{ij}z +\epsilon_{ij}
\end{equation}

In the above equation, $y$ is the outcome variable, $X$ is the matrix of predictors which is the binary variable of each perturbation, $\beta$ is the prediction coefficients (similar to a simple linear regression), $b$ is the matrix of random effects which is based on the number of participants, and $z$ is the coefficients for each random effect (each participant). Lastly, $\epsilon$ is the model error term. 

Also, the elements of the $b$ and $\epsilon$ matrices are defined as follows:

\begin{equation}
    b_{ij} \sim N(0,\psi_k^{2}),Cov(b_k,b_{k'})
\end{equation}

\begin{equation}
    \epsilon_{ij} \sim N(0,\sigma^{2}\lambda_{ijj}),Cov(\epsilon_{ij},\epsilon_{ij'})
\end{equation}

\section*{Acknowledgment}
We would like to thank the UVA Link Lab, and Commonwealth Cyber Initiative (CCI) for providing support and resources to enable this project. Also, we are thankful to the UVA Institutional Review Board for their continuous support and feedback.



\bibliographystyle{IEEEtran}
\bibliography{main}

\begin{thebibliography}{10}
\providecommand{\url}[1]{#1}
\csname url@samestyle\endcsname
\providecommand{\newblock}{\relax}
\providecommand{\bibinfo}[2]{#2}
\providecommand{\BIBentrySTDinterwordspacing}{\spaceskip=0pt\relax}
\providecommand{\BIBentryALTinterwordstretchfactor}{4}
\providecommand{\BIBentryALTinterwordspacing}{\spaceskip=\fontdimen2\font plus
\BIBentryALTinterwordstretchfactor\fontdimen3\font minus
  \fontdimen4\font\relax}
\providecommand{\BIBforeignlanguage}[2]{{%
\expandafter\ifx\csname l@#1\endcsname\relax
\typeout{** WARNING: IEEEtran.bst: No hyphenation pattern has been}%
\typeout{** loaded for the language `#1'. Using the pattern for}%
\typeout{** the default language instead.}%
\else
\language=\csname l@#1\endcsname
\fi
#2}}
\providecommand{\BIBdecl}{\relax}
\BIBdecl

\bibitem{bustos2021predicting}
C.~Bustos, N.~Elhaouij, A.~Sol{\'e}-Ribalta, J.~Borge-Holthoefer, A.~Lapedriza,
  and R.~Picard, ``Predicting driver self-reported stress by analyzing the road
  scene,'' in \emph{2021 9th International Conference on Affective Computing
  and Intelligent Interaction (ACII)}.\hskip 1em plus 0.5em minus 0.4em\relax
  IEEE, 2021, pp. 1--8.

\bibitem{taubman2012effects}
O.~Taubman-Ben-Ari, ``The effects of positive emotion priming on self-reported
  reckless driving,'' \emph{Accident Analysis \& Prevention}, vol.~45, pp.
  718--725, 2012.

\bibitem{roidl2014emotional}
E.~Roidl, B.~Frehse, and R.~H{\"o}ger, ``Emotional states of drivers and the
  impact on speed, acceleration and traffic violations—a simulator study,''
  \emph{Accident Analysis \& Prevention}, vol.~70, pp. 282--292, 2014.

\bibitem{dingus2016driver}
T.~A. Dingus, F.~Guo, S.~Lee, J.~F. Antin, M.~Perez, M.~Buchanan-King, and
  J.~Hankey, ``Driver crash risk factors and prevalence evaluation using
  naturalistic driving data,'' \emph{Proceedings of the National Academy of
  Sciences}, vol. 113, no.~10, pp. 2636--2641, 2016.

\bibitem{sani2017aggression}
S.~R.~H. Sani, Z.~Tabibi, J.~S. Fadardi, and D.~Stavrinos, ``Aggression,
  emotional self-regulation, attentional bias, and cognitive inhibition predict
  risky driving behavior,'' \emph{Accident Analysis \& Prevention}, vol. 109,
  pp. 78--88, 2017.

\bibitem{shukri2022theory}
M.~Shukri, F.~Jones, and M.~Conner, ``Theory of planned behaviour,
  psychological stressors and intention to avoid violating traffic rules: A
  multi-level modelling analysis,'' \emph{Accident Analysis \& Prevention},
  vol. 169, p. 106624, 2022.

\bibitem{zepf2019towards}
S.~Zepf, M.~Dittrich, J.~Hernandez, and A.~Schmitt, ``Towards empathetic car
  interfaces: Emotional triggers while driving,'' in \emph{Extended Abstracts
  of the 2019 CHI Conference on Human Factors in Computing Systems}, 2019, pp.
  1--6.

\bibitem{dittrich2021drivers}
M.~Dittrich, ``Why drivers feel the way they do: An on-the-road study using
  self-reports and geo-tagging,'' in \emph{13th International Conference on
  Automotive User Interfaces and Interactive Vehicular Applications}, 2021, pp.
  116--125.

\bibitem{nacpil2021application}
E.~J.~C. Nacpil, Z.~Wang, and K.~Nakano, ``Application of physiological sensors
  for personalization in semi-autonomous driving: A review,'' \emph{IEEE
  Sensors Journal}, 2021.

\bibitem{zheng2015biosignal}
R.~Zheng, S.~Yamabe, K.~Nakano, and Y.~Suda, ``Biosignal analysis to assess
  mental stress in automatic driving of trucks: Palmar perspiration and
  masseter electromyography,'' \emph{Sensors}, vol.~15, no.~3, pp. 5136--5150,
  2015.

\bibitem{carfollow}
V.~Radhakrishnan, N.~Merat, T.~Louw, R.~Gonçalves, G.~Torrao, W.~Lv,
  P.~Puente~Guillen, and M.~Lenné, ``Physiological indicators of driver
  workload during car-following scenarios and takeovers in highly automated
  driving,'' \emph{Transportation Research Part F Traffic Psychology and
  Behaviour}, vol.~87, pp. 149--163, 05 2022.

\bibitem{tavakoli2021leveraging}
A.~Tavakoli, M.~Boukhechba, and A.~Heydarian, ``Leveraging ubiquitous computing
  for empathetic routing: A naturalistic data-driven approach,'' in
  \emph{Extended Abstracts of the 2021 CHI Conference on Human Factors in
  Computing Systems}, 2021, pp. 1--6.

\bibitem{fakhrhosseini2019angry}
S.~M. FakhrHosseini and M.~Jeon, ``How do angry drivers respond to emotional
  music? a comprehensive perspective on assessing emotion,'' \emph{Journal on
  multimodal user interfaces}, vol.~13, no.~2, pp. 137--150, 2019.

\bibitem{niu2020music}
J.~Niu, C.~Ma, J.~Liu, L.~Li, T.~Hu, and L.~Ran, ``Is music a mediator
  impacting car following when driver’s personalities are considered,''
  \emph{Accident Analysis \& Prevention}, vol. 147, p. 105774, 2020.

\bibitem{wang2021personalized}
Z.~Wang, H.~Xiong, J.~Zhang, S.~Yang, M.~Boukhechba, L.~E. Barnes, D.~Zhang,
  and D.~Dou, ``From personalized medicine to population health: A survey of
  mhealth sensing techniques,'' \emph{arXiv preprint arXiv:2107.00948}, 2021.

\bibitem{tavakoli2021harmony}
A.~Tavakoli, S.~Kumar, X.~Guo, V.~Balali, M.~Boukhechba, and A.~Heydarian,
  ``Harmony: A human-centered multimodal driving study in the wild,''
  \emph{IEEE Access}, vol.~9, pp. 23\,956--23\,978, 2021.

\bibitem{napoletano2018combining}
P.~Napoletano and S.~Rossi, ``Combining heart and breathing rate for car driver
  stress recognition,'' in \emph{2018 IEEE 8th International Conference on
  Consumer Electronics-Berlin (ICCE-Berlin)}.\hskip 1em plus 0.5em minus
  0.4em\relax IEEE, 2018, pp. 1--5.

\bibitem{chung2019methods}
W.-Y. Chung, T.-W. Chong, and B.-G. Lee, ``Methods to detect and reduce driver
  stress: a review,'' \emph{International journal of automotive technology},
  vol.~20, no.~5, pp. 1051--1063, 2019.

\bibitem{lohani2019review}
M.~Lohani, B.~R. Payne, and D.~L. Strayer, ``A review of psychophysiological
  measures to assess cognitive states in real-world driving,'' \emph{Frontiers
  in human neuroscience}, vol.~13, p.~57, 2019.

\bibitem{laora2022}
L.~Kerautret and J.~Navarro, ``Detecting driver stress and hazard anticipation
  using real‐time cardiac measurement: A simulator study,'' \emph{Brain and
  Behavior}, 01 2022.

\bibitem{matterport_maskrcnn_2017}
W.~Abdulla, ``Mask r-cnn for object detection and instance segmentation on
  keras and tensorflow,'' \url{https://github.com/matterport/Mask_RCNN}, 2017.

\bibitem{lin2014microsoft}
T.-Y. Lin, M.~Maire, S.~Belongie, J.~Hays, P.~Perona, D.~Ramanan,
  P.~Doll{\'a}r, and C.~L. Zitnick, ``Microsoft coco: Common objects in
  context,'' in \emph{European conference on computer vision}.\hskip 1em plus
  0.5em minus 0.4em\relax Springer, 2014, pp. 740--755.

\bibitem{barry1993bayesian}
D.~Barry and J.~A. Hartigan, ``A bayesian analysis for change point problems,''
  \emph{Journal of the American Statistical Association}, vol.~88, no. 421, pp.
  309--319, 1993.

\bibitem{izard2009emotion}
C.~E. Izard, ``Emotion theory and research: Highlights, unanswered questions,
  and emerging issues,'' \emph{Annual review of psychology}, vol.~60, pp.
  1--25, 2009.

\bibitem{tracy2011four}
J.~L. Tracy and D.~Randles, ``Four models of basic emotions: a review of ekman
  and cordaro, izard, levenson, and panksepp and watt,'' \emph{Emotion review},
  vol.~3, no.~4, pp. 397--405, 2011.

\bibitem{ekman2011meant}
P.~Ekman and D.~Cordaro, ``What is meant by calling emotions basic,''
  \emph{Emotion review}, vol.~3, no.~4, pp. 364--370, 2011.

\bibitem{barnard2016anxiety}
M.~P. Barnard and P.~Chapman, ``Are anxiety and fear separable emotions in
  driving? a laboratory study of behavioural and physiological responses to
  different driving environments,'' \emph{Accident Analysis \& Prevention},
  vol.~86, pp. 99--107, 2016.

\bibitem{russell1980circumplex}
J.~A. Russell, ``A circumplex model of affect.'' \emph{Journal of personality
  and social psychology}, vol.~39, no.~6, p. 1161, 1980.

\bibitem{francis2018embodied}
A.~L. Francis, ``The embodied theory of stress: A constructionist perspective
  on the experience of stress,'' \emph{Review of General Psychology}, vol.~22,
  no.~4, pp. 398--405, 2018.

\bibitem{chesnut2021stress}
M.~Chesnut, S.~Harati, P.~Paredes, Y.~Khan, A.~Foudeh, J.~Kim, Z.~Bao, and
  L.~M. Williams, ``Stress markers for mental states and biotypes of depression
  and anxiety: A scoping review and preliminary illustrative analysis,''
  \emph{Chronic Stress}, vol.~5, p. 24705470211000338, 2021.

\bibitem{giannakakis2017stress}
G.~Giannakakis, M.~Pediaditis, D.~Manousos, E.~Kazantzaki, F.~Chiarugi, P.~G.
  Simos, K.~Marias, and M.~Tsiknakis, ``Stress and anxiety detection using
  facial cues from videos,'' \emph{Biomedical Signal Processing and Control},
  vol.~31, pp. 89--101, 2017.

\bibitem{kim2018stress}
H.-G. Kim, E.-J. Cheon, D.-S. Bai, Y.~H. Lee, and B.-H. Koo, ``Stress and heart
  rate variability: A meta-analysis and review of the literature,''
  \emph{Psychiatry investigation}, vol.~15, no.~3, p. 235, 2018.

\bibitem{tavakoli2020personalized}
A.~Tavakoli, M.~Boukhechba, and A.~Heydarian, ``Personalized driver state
  profiles: A naturalistic data-driven study,'' in \emph{International
  Conference on Applied Human Factors and Ergonomics}.\hskip 1em plus 0.5em
  minus 0.4em\relax Springer, 2020, pp. 32--39.

\bibitem{du2020psychophysiological}
N.~Du, X.~J. Yang, and F.~Zhou, ``Psychophysiological responses to takeover
  requests in conditionally automated driving,'' \emph{arXiv preprint
  arXiv:2010.03047}, 2020.

\bibitem{mcduff2016affdex}
D.~McDuff, A.~Mahmoud, M.~Mavadati, M.~Amr, J.~Turcot, and R.~e. Kaliouby,
  ``Affdex sdk: a cross-platform real-time multi-face expression recognition
  toolkit,'' in \emph{Proceedings of the 2016 CHI conference extended abstracts
  on human factors in computing systems}, 2016, pp. 3723--3726.

\bibitem{scott2018qualitative}
B.~Scott-Parker, C.~M. Jones, K.~Rune, and J.~Tucker, ``A qualitative
  exploration of driving stress and driving discourtesy,'' \emph{Accident
  Analysis \& Prevention}, vol. 118, pp. 38--53, 2018.

\bibitem{mesken2007frequency}
J.~Mesken, M.~P. Hagenzieker, T.~Rothengatter, and D.~De~Waard, ``Frequency,
  determinants, and consequences of different drivers’ emotions: An
  on-the-road study using self-reports,(observed) behaviour, and physiology,''
  \emph{Transportation research part F: traffic psychology and behaviour},
  vol.~10, no.~6, pp. 458--475, 2007.

\bibitem{roidl2013emotional}
E.~Roidl, B.~Frehse, M.~Oehl, and R.~H{\"o}ger, ``The emotional spectrum in
  traffic situations: Results of two online-studies,'' \emph{Transportation
  research part F: traffic psychology and behaviour}, vol.~18, pp. 168--188,
  2013.

\bibitem{tavakoli2022multimodal}
A.~Tavakoli and A.~Heydarian, ``Multimodal driver state modeling through
  unsupervised learning,'' \emph{Accident Analysis \& Prevention}, vol. 170, p.
  106640, 2022.

\bibitem{Harmonydata}
\BIBentryALTinterwordspacing
UVABRAINLAB. (2020) Harmony case study. [Online]. Available:
  \url{https://osf.io/zextd/}
\BIBentrySTDinterwordspacing

\bibitem{he2017mask}
K.~He, G.~Gkioxari, P.~Doll{\'a}r, and R.~B. Girshick, ``Mask r-cnn. corr
  abs/1703.06870 (2017),'' \emph{arXiv preprint arXiv:1703.06870}, 2017.

\bibitem{githubLai}
\BIBentryALTinterwordspacing
N.~Lai and A.~Tavakoli. (2022) Code for stop sign detection. [Online].
  Available: \url{https://github.com/nathan-lai-engineering/}
\BIBentrySTDinterwordspacing

\bibitem{redmon2015you}
J.~Redmon, S.~Divvala, R.~Girshick, and A.~Farhadi, ``You only look once:
  Unified, real-time object detection. arxiv 2015,'' \emph{arXiv preprint
  arXiv:1506.02640}, 2015.

\bibitem{jocher2020yolov5}
G.~Jocher, K.~Nishimura, T.~Mineeva, and R.~Vilari{\~n}o, ``Yolov5,''
  \emph{Code repository https://github. com/ultralytics/yolov5}, 2020.

\bibitem{glenn_jocher_2021_5563715}
\BIBentryALTinterwordspacing
G.~J. et. al., ``{ultralytics/yolov5: v6.0 - YOLOv5n 'Nano' models, Roboflow
  integration, TensorFlow export, OpenCV DNN support},'' Oct. 2021. [Online].
  Available: \url{https://doi.org/10.5281/zenodo.5563715}
\BIBentrySTDinterwordspacing

\bibitem{jensen2016vision}
M.~B. Jensen, M.~P. Philipsen, A.~M{\o}gelmose, T.~B. Moeslund, and M.~M.
  Trivedi, ``Vision for looking at traffic lights: Issues, survey, and
  perspectives,'' \emph{IEEE Transactions on Intelligent Transportation
  Systems}, vol.~17, no.~7, pp. 1800--1815, 2016.

\bibitem{balali2016evaluation}
V.~Balali and M.~Golparvar-Fard, ``Evaluation of multiclass traffic sign
  detection and classification methods for us roadway asset inventory
  management,'' \emph{Journal of Computing in Civil Engineering}, vol.~30,
  no.~2, p. 04015022, 2016.

\bibitem{nguyen2021survey}
T.~T.~H. Nguyen, A.~Jatowt, M.~Coustaty, and A.~Doucet, ``Survey of post-ocr
  processing approaches,'' \emph{ACM Computing Surveys (CSUR)}, vol.~54, no.~6,
  pp. 1--37, 2021.

\bibitem{pytesseract}
\BIBentryALTinterwordspacing
(2021, Sep) Python tesseract. [Online]. Available:
  \url{https://pypi.org/project/pytesseract/}
\BIBentrySTDinterwordspacing

\bibitem{EasyOCR}
\BIBentryALTinterwordspacing
(2021, Sep) Easy ocr. [Online]. Available:
  \url{https://github.com/JaidedAI/EasyOCR}
\BIBentrySTDinterwordspacing

\bibitem{malladi2013online}
R.~Malladi, G.~P. Kalamangalam, and B.~Aazhang, ``Online bayesian change point
  detection algorithms for segmentation of epileptic activity,'' in \emph{2013
  Asilomar Conference on Signals, Systems and Computers}.\hskip 1em plus 0.5em
  minus 0.4em\relax IEEE, 2013, pp. 1833--1837.

\bibitem{guo2021benchmarking}
X.~Guo, E.~M. Robartes, A.~Angulo, T.~D. Chen, and A.~Heydarian, ``Benchmarking
  the use of immersive virtual bike simulators for understanding cyclist
  behaviors,'' 2021.

\bibitem{guo2021orclsim}
X.~Guo, A.~Angulo, E.~Robartes, T.~D. Chen, and A.~Heydarian, ``Orclsim: A
  system architecture for studying bicyclist and pedestrian physiological
  behavior through immersive virtual environments,'' \emph{arXiv preprint
  arXiv:2112.03420}, 2021.

\bibitem{kumarlever}
S.~Kumar, D.~Datta, G.~Dong, L.~Cai, L.~Barnes, and M.~Boukhechba, ``Leveraging
  mobile sensing and bayesian change point analysis to monitor community-scale
  behavioral interventions: a case study on covid-19,'' 12 2021.

\bibitem{dong2021detection}
G.~Dong, L.~Cai, S.~Kumar, D.~Datta, L.~E. Barnes, and M.~Boukhechba,
  ``Detection and analysis of interrupted behaviors by public policy
  interventions during covid-19,'' in \emph{2021 IEEE/ACM Conference on
  Connected Health: Applications, Systems and Engineering Technologies
  (CHASE)}.\hskip 1em plus 0.5em minus 0.4em\relax IEEE, 2021, pp. 46--57.

\bibitem{erdman2007bcp}
C.~Erdman and J.~W. Emerson, ``bcp: an r package for performing a bayesian
  analysis of change point problems,'' \emph{Journal of Statistical Software},
  vol.~23, no.~1, pp. 1--13, 2007.

\bibitem{brown2021introduction}
V.~A. Brown, ``An introduction to linear mixed-effects modeling in r,''
  \emph{Advances in Methods and Practices in Psychological Science}, vol.~4,
  no.~1, p. 2515245920960351, 2021.

\bibitem{fox2002linear}
J.~Fox, ``Linear mixed models,'' \emph{Appendix to an R and S-plus Companion to
  Applied Regression}, vol.~16, pp. 2349--2380, 2002.

\bibitem{bolker2009generalized}
B.~M. Bolker, M.~E. Brooks, C.~J. Clark, S.~W. Geange, J.~R. Poulsen, M.~H.~H.
  Stevens, and J.-S.~S. White, ``Generalized linear mixed models: a practical
  guide for ecology and evolution,'' \emph{Trends in ecology \& evolution},
  vol.~24, no.~3, pp. 127--135, 2009.

\bibitem{bates2007lme4}
D.~Bates, D.~Sarkar, M.~D. Bates, and L.~Matrix, ``The lme4 package,'' \emph{R
  package version}, vol.~2, no.~1, p.~74, 2007.

\bibitem{R_lang}
\BIBentryALTinterwordspacing
{R Core Team}, \emph{R: A Language and Environment for Statistical Computing},
  R Foundation for Statistical Computing, Vienna, Austria, 2017. [Online].
  Available: \url{https://www.R-project.org/}
\BIBentrySTDinterwordspacing

\bibitem{mcduff2013affectiva}
D.~McDuff, R.~Kaliouby, T.~Senechal, M.~Amr, J.~Cohn, and R.~Picard,
  ``Affectiva-mit facial expression dataset (am-fed): Naturalistic and
  spontaneous facial expressions collected,'' in \emph{Proceedings of the IEEE
  Conference on Computer Vision and Pattern Recognition Workshops}, 2013, pp.
  881--888.

\bibitem{kulke2020comparison}
L.~Kulke, D.~Feyerabend, and A.~Schacht, ``A comparison of the affectiva
  imotions facial expression analysis software with emg for identifying facial
  expressions of emotion,'' \emph{Frontiers in psychology}, vol.~11, p. 329,
  2020.

\bibitem{abdic2016driver}
I.~Abdic, L.~Fridman, D.~McDuff, E.~Marchi, B.~Reimer, and B.~Schuller,
  ``Driver frustration detection from audio and video in the wild,'' in
  \emph{Springer}, vol. 9904.\hskip 1em plus 0.5em minus 0.4em\relax Springer,
  2016, p. 237.

\bibitem{mehta2021self}
A.~Mehta, C.~Sharma, M.~Kanala, M.~Thakur, R.~Harrison, and D.~D. Torrico,
  ``Self-reported emotions and facial expressions on consumer acceptability: A
  study using energy drinks,'' \emph{Foods}, vol.~10, no.~2, p. 330, 2021.

\bibitem{reinares2019cognitive}
P.~Reinares-Lara, A.~Rodr{\'\i}guez-Fuertes, and B.~Garcia-Henche, ``The
  cognitive dimension and the affective dimension in the patient’s
  experience,'' \emph{Frontiers in psychology}, vol.~10, p. 2177, 2019.

\bibitem{tavakoli2019multimodal}
A.~Tavakoli, V.~Balali, and A.~Heydarian, ``A multimodal approach for
  monitoring driving behavior and emotions,'' Transportation Research Board,
  Washington DC, United States, Tech. Rep., 2019.

\bibitem{wu2021yolop}
D.~Wu, M.~Liao, W.~Zhang, and X.~Wang, ``Yolop: You only look once for panoptic
  driving perception,'' \emph{arXiv preprint arXiv:2108.11250}, 2021.

\bibitem{chawla2002smote}
N.~V. Chawla, K.~W. Bowyer, L.~O. Hall, and W.~P. Kegelmeyer, ``Smote:
  synthetic minority over-sampling technique,'' \emph{Journal of artificial
  intelligence research}, vol.~16, pp. 321--357, 2002.

\bibitem{mckight2010kruskal}
P.~E. McKight and J.~Najab, ``Kruskal-wallis test,'' \emph{The corsini
  encyclopedia of psychology}, pp. 1--1, 2010.

\bibitem{holm1979simple}
S.~Holm, ``A simple sequentially rejective multiple test procedure,''
  \emph{Scandinavian journal of statistics}, pp. 65--70, 1979.

\bibitem{muller2007dynamic}
M.~M{\"u}ller, ``Dynamic time warping,'' \emph{Information retrieval for music
  and motion}, pp. 69--84, 2007.

\bibitem{berndt1994using}
D.~J. Berndt and J.~Clifford, ``Using dynamic time warping to find patterns in
  time series.'' in \emph{KDD workshop}, vol.~10, no.~16.\hskip 1em plus 0.5em
  minus 0.4em\relax Seattle, WA, USA:, 1994, pp. 359--370.

\bibitem{JMLR:v21:20-091}
\BIBentryALTinterwordspacing
R.~Tavenard, J.~Faouzi, G.~Vandewiele, F.~Divo, G.~Androz, C.~Holtz, M.~Payne,
  R.~Yurchak, M.~Ru{\ss}wurm, K.~Kolar, and E.~Woods, ``Tslearn, a machine
  learning toolkit for time series data,'' \emph{Journal of Machine Learning
  Research}, vol.~21, no. 118, pp. 1--6, 2020. [Online]. Available:
  \url{http://jmlr.org/papers/v21/20-091.html}
\BIBentrySTDinterwordspacing

\bibitem{jain1999data}
A.~K. Jain, M.~N. Murty, and P.~J. Flynn, ``Data clustering: a review,''
  \emph{ACM computing surveys (CSUR)}, vol.~31, no.~3, pp. 264--323, 1999.

\bibitem{shahapure2020cluster}
K.~R. Shahapure and C.~Nicholas, ``Cluster quality analysis using silhouette
  score,'' in \emph{2020 IEEE 7th International Conference on Data Science and
  Advanced Analytics (DSAA)}.\hskip 1em plus 0.5em minus 0.4em\relax IEEE,
  2020, pp. 747--748.

\bibitem{wagenmakers2004aic}
E.-J. Wagenmakers and S.~Farrell, ``Aic model selection using akaike weights,''
  \emph{Psychonomic bulletin \& review}, vol.~11, no.~1, pp. 192--196, 2004.

\bibitem{radhakrishnan2022physiological}
V.~Radhakrishnan, N.~Merat, T.~Louw, R.~C. Gon{\c{c}}alves, G.~Torrao, W.~Lyu,
  P.~P. Guillen, and M.~G. Lenn{\'e}, ``Physiological indicators of driver
  workload during car-following scenarios and takeovers in highly automated
  driving,'' \emph{Transportation Research Part F: Traffic Psychology and
  Behaviour}, vol.~87, pp. 149--163, 2022.

\bibitem{milardo2021understanding}
S.~Milardo, P.~Rathore, M.~Amorim, U.~Fugiglando, P.~Santi, and C.~Ratti,
  ``Understanding drivers' stress and interactions with vehicle systems through
  naturalistic data analysis,'' \emph{IEEE Transactions on Intelligent
  Transportation Systems}, 2021.

\bibitem{tavakoli2021statespace}
A.~Tavakoli, S.~Boker, and A.~Heydarian, ``Driver state modeling through latent
  variable state space framework in the wild,'' \emph{arXiv preprint
  arXiv:2203.00834}, 2022.

\end{thebibliography}

\begin{IEEEbiography}[{\includegraphics[width=1in,height=1.25in,clip,keepaspectratio]{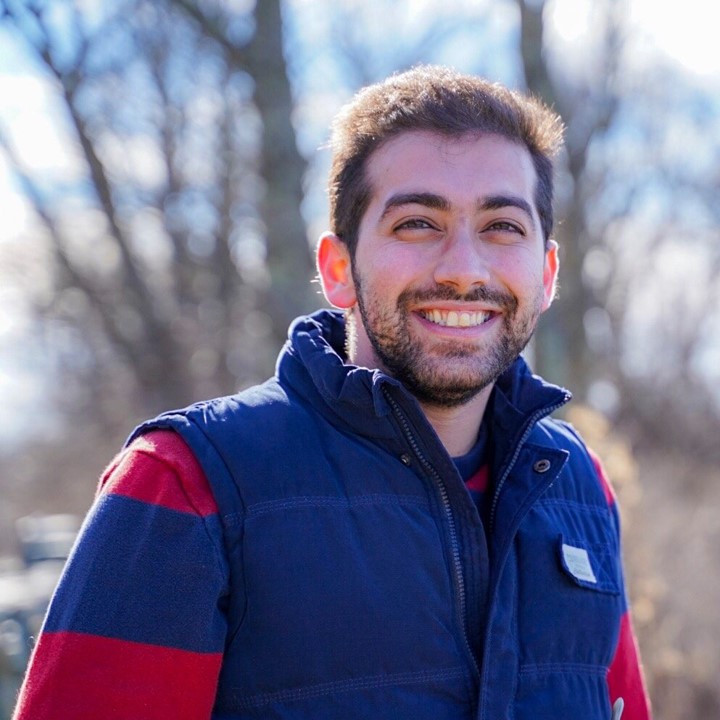}}]{Arash Tavakoli} Arash Tavakoli is a PhD student in the Engineering Systems and Environment department as well as the Link Lab at the University of Virginia. He has earned his BSc. and MSc. in Civil Engineering from the Sharif University of Technology and Virginia Tech, respectively. Arash’s research interest lies on the intersection of transportation engineering, computer science, and psychology.
\end{IEEEbiography}

\begin{IEEEbiography}[{\includegraphics[width=1in,height=1.25in,clip,keepaspectratio]{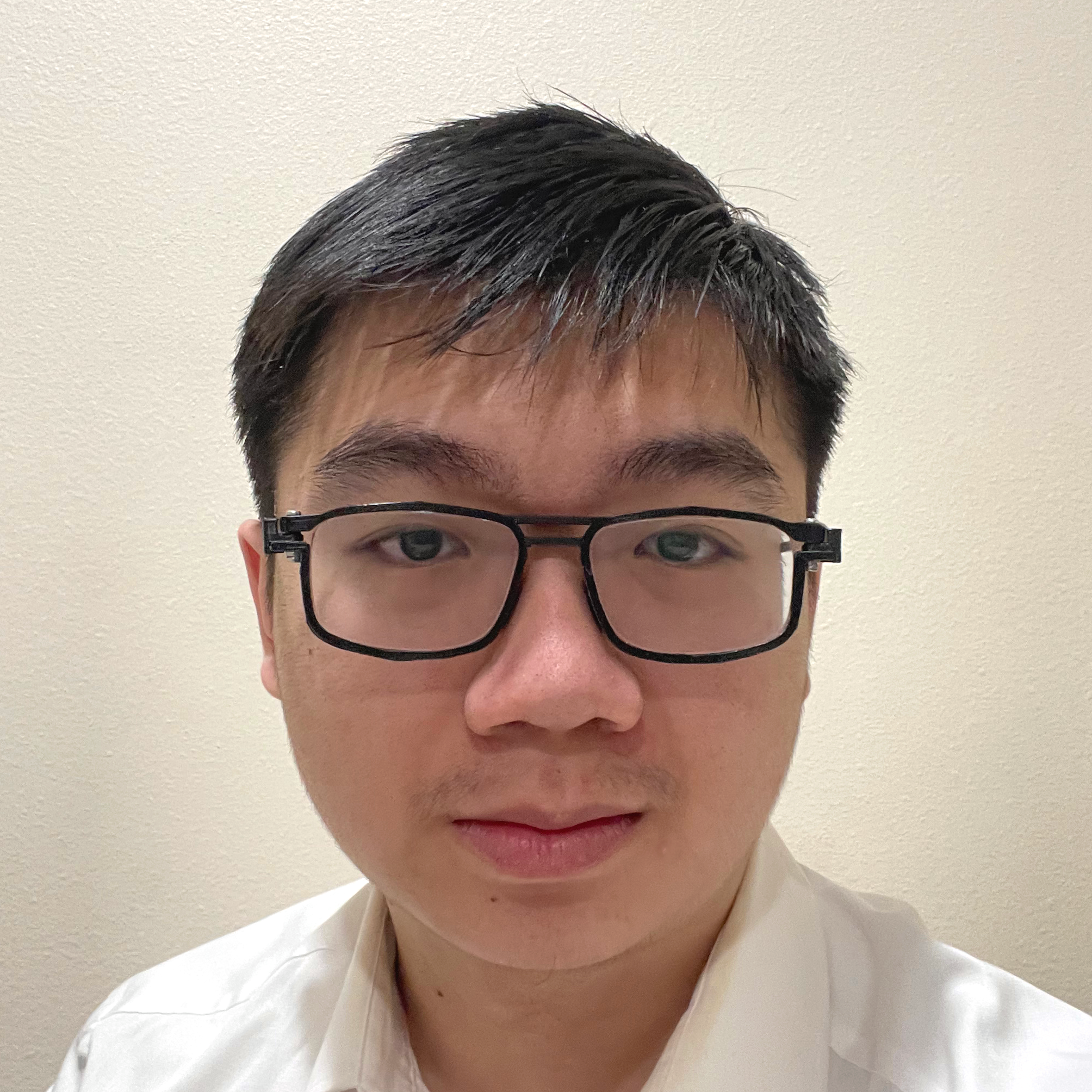}}]{Nathan Lai} Nathan Lai is an undergraduate student in his 3rd year of obtaining a BSc. in Computer Science at California State University, Long Beach. Nathan’s research interests include computer vision and machine learning. 
\end{IEEEbiography}

\begin{IEEEbiography}[{\includegraphics[width=1in,height=1.25in,clip,keepaspectratio]{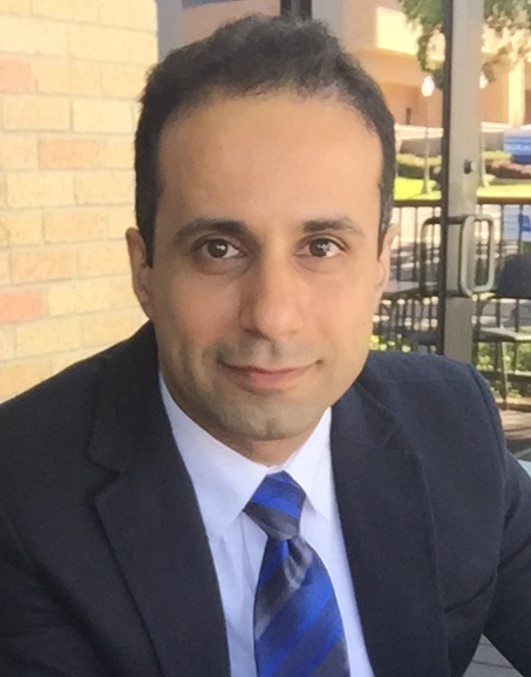}}]{Vahid Balali} Dr. Vahid Balali is an Associate Professor in the Department of Civil Engineering and Construction Engineering Management at CSULB. His research focuses on the (1) visual data sensing and analytics for the AEC industry, (2) virtual design and construction for infrastructure asset management and interoperable system integration, and (3) smart cities in transportation for sustainable infrastructure decision-making. He is currently an associate member of ASCE and CMAA, committee member of the ASCE Data Sensing and Analysis (DSA) and ASCE Visual Information Modeling and Simulation (VIMS) committees, and friend member of relevant TRB committees. He received his Ph.D. in Civil Engineering from the University of Illinois at Urbana-Champaign (UIUC), his second M.Sc. in Construction Engineering and Management from Virginia Tech, and his M.Sc. and B.Sc. in Civil Engineering from the University of Tehran.
\end{IEEEbiography}

\begin{IEEEbiography}[{\includegraphics[width=1in,height=1.25in,clip,keepaspectratio]{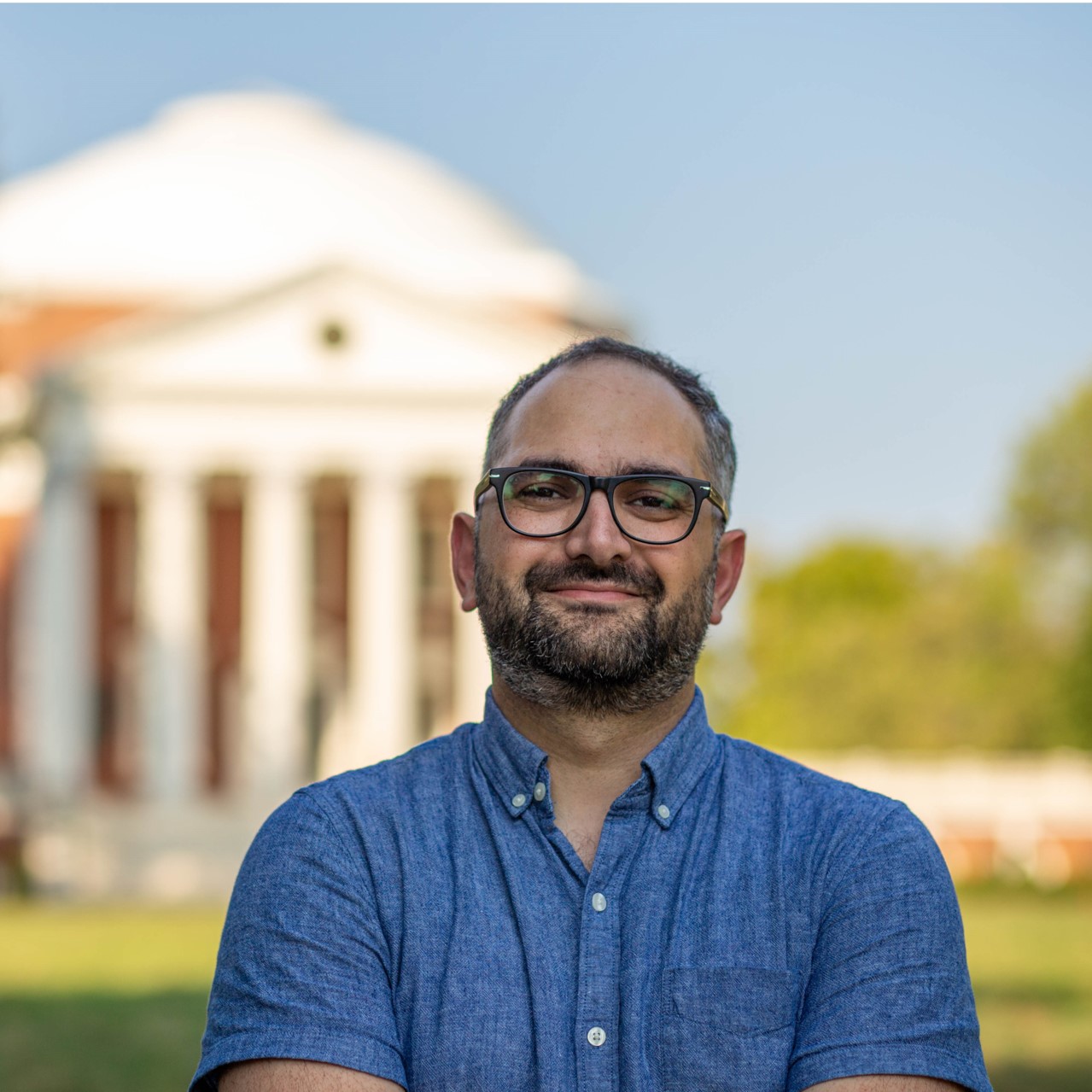}}]{Arsalan Heydarian} Arsalan Heydarian is an Assistant Professor in the department of Engineering Systems and Environment as well as the UVA LINK LAB. His research focuses on user-centered design, construction, and operation of intelligent infrastructure with the objective of enhancing sustainability, adaptability, and resilience future infrastructure systems. Specifically, his research can be divided into four main research streams: (1) intelligent built environments; (2) mobility and infrastructure design; (3) smart transportation and (4) data-driven mixed reality. Dr. Heydarian received his Ph.D. in Civil Engineering from the University of Southern California (USC), M.Sc in System Engineering from USC, and B.Sc. and M.Sc in Civil Engineering from Virginia Tech. 
\end{IEEEbiography}

\end{document}